\documentclass{article}
\usepackage[utf8]{inputenc}

\usepackage{graphicx,color}

\usepackage{jcappub} 
\usepackage{url}
\usepackage[T1]{fontenc} 

\def\iso#1#2{\mbox{${}^{#2}{\rm #1}$}}
\def\he#1{\iso{He}{#1}}
\def\li#1{\iso{Li}{#1}}

\def\mpl{M_{\rm Pl}}

\def\pfrac#1#2{\left( \frac{#1}{#2} \right)}

\def\fdm{f_{\rm DM}}
\def\fgam{f_{\gamma}}
\def\fgamc{f_{\gamma,0}}

\newcommand{\beq}{\begin{equation}\begin{aligned}}
\newcommand{\eeq}{\end{aligned}\end{equation}}
\newcommand{\Neff}{N_{\mathrm{eff}}}

\begin{document}

\begin{flushright}
UMN--TH--4125/22, FTPI--MINN--22/16   \\
July 2022
\end{flushright}

\title{\boldmath Probing Physics Beyond the Standard Model:  Limits from BBN and the CMB Independently and Combined}


\author[a,1]{Tsung-Han Yeh,\note{Corresponding author.}} 
\author[a,b]{Jessie Shelton,} 
\author[c]{Keith A. Olive,} 
\author[a,b,d]{Brian D. Fields} 


\affiliation[a]{Department of Physics, University of Illinois, Urbana IL 61801}
\affiliation[b]{Illinois Center for Advanced Studies of the Universe}
\affiliation[c]{William I. Fine Theoretical Physics Institute, School of
 Physics and Astronomy, University of Minnesota, Minneapolis, MN 55455}
\affiliation[d]{Department of Astronomy, University of Illinois, Urbana, IL 61801}

\emailAdd{tyeh6@illinois.edu}

\abstract{We present new Big Bang Nucleosynthesis (BBN) limits on the cosmic expansion rate or relativistic energy density, quantified via the number $N_\nu$ of equivalent neutrino species.  
We use the latest light element observations, neutron mean lifetime, 
and update our evaluation for the nuclear rates $d+d \rightarrow \he3 + n$ and $d+d \rightarrow \iso{H}{3}+ p$. 
Combining this result with the independent constraints from the cosmic microwave background (CMB) 
yields tight limits on new physics that perturbs $N_\nu$ and $\eta$ prior to 
cosmic nucleosynthesis: 
a joint BBN+CMB analysis gives $N_\nu = 2.898 \pm 0.141$,
resulting in $N_\nu < 3.180$ at $2\sigma$.
We apply these limits to a wide variety of 
new physics scenarios including right-handed neutrinos, dark radiation, and a stochastic gravitational wave background.
The strength of the independent BBN and CMB constraints
now opens a new window: we can search for limits on 
potential {\em changes} in $N_\nu$ and/or the baryon-to-photon ratio $\eta$ between 
the two epochs.
The present data
place strong constraints on the allowed changes in $N_\nu$ between BBN and CMB decoupling; for example, we find $-0.708 < N_\nu^{\rm CMB}-N_\nu^{\rm BBN} < 0.328$
in the case where $\eta$ and the primordial helium mass fraction $Y_p$ are unchanged between the two epochs; we also give
limits on the allowed variations in $\eta$ or
in $(\eta,N_\nu)$ jointly.
We discuss scenarios in which such changes could occur, and show that BBN+CMB results combine to place important constraints on some early dark energy
models to explain the $H_0$ tension.
Looking to the future, we forecast the tightened precision for $N_\nu$ arising from
both CMB Stage 4 measurements as well as improvements in astronomical \he4 measurements.  We find that CMB-S4 combined with present BBN and light element observation precision can give $\sigma(N_\nu) \simeq 0.03$.  Such future precision would reveal the expected effect of neutrino heating ($N_{\rm eff}-3=0.044$) of the CMB during BBN, and would be near the level to reveal any particle species ever in thermal equilibrium with the standard model. Improved $Y_p$ measurements can push this precision even further.}

\maketitle

\flushbottom

\section{Introduction}\label{intro}

Big-bang nucleosynthesis (BBN) is the theory which accounts for the production
of the lightest nuclei~\cite{Alpher:1949sef} during the first seconds to minutes
of cosmic time (for reviews see refs.~\cite{1991ApJ...376...51W,2000PhR...333..389O,Steigman2007,Iocco2009,Cyburt2016,Pitrou2018}).  The physics behind standard BBN 
is very well known because the typical energy scale is that of relatively low-energy nuclear physics, i.e., of order 1 MeV. We refer to 
standard BBN (SBBN) as the theory based on the Standard Model of particle and nuclear interactions, and $\Lambda$CDM cosmology with three neutrino flavors. At present, SBBN
is the 
earliest reliable probe the universe where the microphysics
is well understood.\footnote{The cosmic microwave background (CMB) probes the early universe to eV energy scales, and as we discuss is an important tool in BBN analysis \cite{Planck:2018vyg}. The CMB in principle also probes back to inflationary times through determinations of the anisotropy spectrum.}
All four fundamental interactions participate in element
formation, so SBBN probes all known interactions.
Moreover, through the effects of the cosmic expansion rate, BBN
is sensitive to all contributions to cosmic mass-energy density
\cite{Hoyle1964,Shvartsman1969,Peebles1971,SSG1977} which can often be parameterized as additional neutrino flavors and as we will see allows one to probe departures from the Standard Model.

In the pioneering work of Steigman, Schramm, and Gunn \cite{SSG1977}, the existence of additional massive charged leptons was assumed to be
accompanied by (massless) neutrinos. As such, these would contribute to the radiation energy density prior to nucleosynthesis,
\beq
\rho_{R} = \rho_\gamma + \rho_e + \rho_\nu =  \frac{\pi^2}{30} \left(2 + \frac72 + \frac74 N_\nu \right) T^4 \, ,
\eeq
for temperatures $T \gtrsim 1$ MeV, where $N_\nu$ is the total number of neutrino flavors (relativistic at temperature $T$).
More generally, the presence of any additional relativistic particle species can be expressed in terms of the equivalent number $N_\nu$.

The best accelerator limit on the number of neutrino species
is based on the invisible width of the $Z^0$ boson.
The LEP experiments combined give $N_\nu = 2.9963 \pm 0.0074$ \cite{Janot:2019oyi};
this of course is in
impressive agreement with the $N_\nu^{\rm SM}=3$ result
from the Standard Model.  Note that this limit applies to
particles that have standard electroweak couplings to the $Z^0$,
whereas the BBN (and CMB) limits probe any species in thermal equilibrium
during this epoch. 
The cosmological limits are thus complementary to the accelerator
results in that they are arise from different physics, and importantly,
these probes can give different results.
We will refer to BBN with $N_\nu \ne 3$ as ``NBBN''.  

As is well known, the abundances of the light elements produced in BBN are very sensitive to the neutron-proton ratio $(n/p)$, when the deuterium bottleneck is broken at $T \sim 0.1$ MeV. While the helium abundance is very sensitive to this ratio (as nearly all free neutrons end up in a \he4 nucleus), the deuterium abundance is also sensitive to $(n/p)$. The neutron-to-proton ratio is largely determined at the freeze-out of the weak interaction rates (modulo neutron
decays). The freeze-out temperature in turn can be determined roughly
by balancing the weak interaction rates with the expansion rate of the Universe, $\Gamma_{\rm weak}(T_f) = H(T_f)$, where $H$ is the Hubble parameter. Furthermore, we can approximate
\begin{eqnarray}
\Gamma_{\rm weak}(T) & \propto & G_F^2 T^5 
\label{Gweak}\\
H^2(T) & \simeq & \frac{8 \pi}{3} G_N \rho_R \, ,
\label{H}
\end{eqnarray}
where $G_{F,N}$ represent Fermi's and Newton's constants. 
Clearly, any change in the number of relativistic degrees of freedom,
will affect $H$ and through $T_f$ will alter $(n/p)_f \sim e^{-\Delta m/T_f}$, where $\Delta m = 1.29$ MeV is the neutron-proton mass difference. It is not hard to convince oneself that an increase in 
$N_\nu$ will lead to a larger value for $(n/p)_f$ and hence a larger value for the the \he4 mass fraction, denoted as $Y_p$.
More generally, any departure from the Standard Model (of either particle/nuclear physics or cosmology) which affects either $\Gamma_{\rm weak}$ or $H$, will alter the light element abundances. 

Of course, constraints on new physics requires accurate abundance measurements, accurate nuclear rates, as well as an accurate determination of the baryon density, $\Omega_b h^2$,
or baryon-to-photon ratio, $\eta$. The baryon density has indeed been determined 
very accurately first by WMAP \cite{WMAP:2003elm} and subsequently by Planck \cite{Planck:2018vyg}, effectively making SBBN a parameter-free theory \cite{2002APh....17...87C}. The deuterium abundance, observed in high redshift quasar absorption systems,  is now determined with approximately 1\% accuracy, \cite{2012MNRAS.425.2477P,2014ApJ...781...31C,2015MNRAS.447.2925R,2016ApJ...830..148C,2016MNRAS.458.2188B,2017MNRAS.468.3239R,2018MNRAS.477.5536Z,2018ApJ...855..102C}
giving
\beq \label{dhobs}
\left(\frac{\rm D}{\rm H}\right)_{\rm obs} = (2.55 \pm 0.03) \times 10^{-5} \, .
\eeq
Because of the small uncertainty in its observational determination, deuterium, which scales as $N_\nu^{0.405}$ \cite{2020JCAP...03..010F,2021Yeh}, now plays an important role in containing physics beyond the Standard Model. Historically however, it is \he4 which
has set the strongest constraints on $N_\nu$.
\he4 is observed in extragalactic HII regions
using a series of \he4 and H emission lines.
The observational determination of \he4
has also improved \cite{2015JCAP...07..011A,2021JCAP...03..027A}.
A recent analysis including high quality observations of the Leoncino dwarf galaxy leads to  an inferred primordial abundance of \citep{aver2021}
\beq
\label{eq:Ypobs}
Y_{p,\rm obs} = 0.2448 \pm 0.0033 \, .
\eeq
Similar recent analyses yield
$0.2453 \pm 0.0034$ \citep{2021JCAP...03..027A},
$0.2436 \pm 0.0040$ \cite{Hsyu:2020uqb}, and
$0.2462 \pm 0.0022$ \cite{Kurichin:2021ppm}.
The helium abundance scales as $Y_p \propto N_\nu^{0.163}$ \cite{2020JCAP...03..010F}. For comparison, using the 
Planck likelihood chains \cite{Planck:2018vyg} with fixed $N_\nu$,
SBBN leads to mean values \cite{2021Yeh} of
\begin{eqnarray}
\left(\frac{\rm D}{\rm H} \right)_{\rm SBBN} & = & (2.506 \pm 0.110) \times 10^{-5} \label{yofdh} \\
Y_{p,\rm SBBN} & = & 0.2469 \pm 0.0002 \, .
\end{eqnarray}

CMB anisotropies, particularly
at high multipoles, are sensitive to the neutrino number
via the effects on the expansion rate, and via
the ratio of the photon diffusion length to the sound horizon
(see, e.g., ref.~\cite{Hou2013}).  In the Standard Model, the effective number of neutrino species $N_{\rm eff}$ is 3.044 \cite{Akita:2020szl,Bennett:2020zkv,EscuderoAbenza:2020cmq,Froustey:2020mcq}. The difference
\beq
\label{eq:nuheat}
\Delta N_{\nu, \ \nu-\rm heating} = N_{\rm eff}-3=0.044
\eeq
is
due to residual heating to neutrinos when accounting for the fact the $e^+ e^-$ annihilation in the early universe is not instantaneous, and has a branching to $e^+ e^-  \rightarrow \nu \bar{\nu}$. In our notation, $N_{\rm eff}^{\rm SM}=3.044$ is equivalent to $N_{\nu}^{\rm SM}=3$ in the standard case.
Eq.~(\ref{eq:nuheat}) thus sets an important target for measurements of $\Delta N_\nu$.

Another target for $N_\nu$ measurements comes from the presence of particles
beyond the Standard Model; here too there is an important physically motivated limit
(see, e.g., the recent review and forecasts in ref. \cite{Dvorkin2022}).
A new species $X$, in equilibrium,  has energy density 
\begin{equation}
    \rho_X = g_{X,\rm eff} \frac{\pi^2}{30} \ T_X^4
\end{equation}
where $g_{X,\rm eff}$ is the number of degrees of freedom for scalars and 7/8 times the number of degrees of freedom when $X$ is a fermion.
At high temperatures, if $X$ is in equilibrium with the SM thermal bath, $T_X = T_\gamma$.  However if $X$ drops out of equilibrium, 
entropy conservation gives $T_X$ at lower temperatures
\beq
\frac{T_X}{T_\gamma} = \left(\frac{g_*(T_{\nu,\rm f})}{g_*(T_{X,\rm f})}\right)^{1/3} \times \left(\frac{4}{11} \right)^{1/3} \, ,
\eeq
where $g_*(T_{\nu,\rm f}) =43/4$ is the number of SM degrees of freedom when neutrinos freeze out and  $g_*(T_{X,\rm f})$ counts the number of effective degrees of freedom (not including $X$) when $X$ freezes out. 
We can relate the density in $X$ to an effective contribution to $N_\nu$ by defining
\beq
\Delta N_\nu = \frac47 g_{X,\rm eff} \left( \frac{T_X}{T_\nu} \right)^{4}
\label{nnutxtnu}
\eeq
recalling that $(T_\nu/T_\gamma)^3 = 4/11$. 
Thus the earlier a new species $X$ freezes out from the Standard Model, the
lower its contribution to $\Delta N_\nu$ due to the resulting dilution of $g_*$:
\begin{eqnarray}
\Delta N_\nu & = & \frac{4}{7} g_{X,\rm eff} \left[ \frac{43}{4 g_*(T_{X,\rm f})} \right]^{4/3} \\
\label{eq:dNlimSM}
& \ge & 0.027 \ g_{X,\rm eff}
\end{eqnarray}
where the limiting value in Eq.~(\ref{eq:dNlimSM}) assumes that the species
freezes out before the entire Standard Model, i.e., $g_*(T_{X,\rm f}) = 427/4$;
later freeze-out will give higher $\Delta N_\nu$. 
We see that this limit is comparable to the neutrino heating 
perturbation in Eq.~(\ref{eq:nuheat}).

Including particles beyond the Standard Model increases the maximum possible $g_*$.
In the extreme case, where new fields decouple so early that there entropy is diluted by 
the field content of a (supersymmetric) grand unified theory, many new (nearly) massless fields are allowed.
For example, in minimal supersymmetric SU(5), $g_*(T_{X,\rm f}) = 1545/4$, the one-sided limit of $\delta N_\nu < 0.226$ (see below)
would place a limit of roughly 47 new scalars or 54 fermionic degrees of freedom. 
This may constrain some string theories which predict large numbers of light moduli \cite{KO}.

In recent years, the cosmic microwave background (CMB)
measurements have also become a probe of $N_\nu$
at the epoch of recombination.  
Allowing $N_\nu$ to vary, the {\it Planck} likelihood chains\footnote{The chains we employ do not assume any relation between the \he4 abundance and the baryon density and as such differ slightly from the values quoted by {\it Planck} in ref.~\cite{Planck:2018vyg}.} \cite{Planck:2018vyg} lead to a determination of $\Omega_{\rm b}^{\rm CMB} h^2 = 0.02224\pm0.00022$ corresponding to $\eta^{\rm CMB} = (6.090\pm0.061) \times 10^{-10}$, and an effective number of neutrino flavors
\beq
N_{\nu}^{\rm CMB} = 2.800 \pm 0.294 \, .
\eeq

When $N_\nu$ is allowed to vary ($N_\nu \ne 3$), the NBBN
calculations of D/H and $Y_p$ are somewhat different \cite{2021Yeh}
\begin{eqnarray}
\left(\frac{\rm D}{\rm H} \right)_{\rm NBBN} & = & (2.447 \pm 0.137) \times 10^{-5} \\
Y_{p,\rm NBBN} & = & 0.2441 \pm 0.0041 \, .
\end{eqnarray}
Note that both are still in very good agreement with observations. Also note that the theory uncertainty in $Y_p$ is now significantly larger due to the strong dependence of $Y_p$ on $N_\nu$. The combined result for NBBN convolved with the CMB chains gives \cite{2021Yeh} $\eta^{\rm NBBN+CMB} = 6.092\pm0.054$ and 
\beq
\left. N_{\nu} \right|_{\rm Yeh2021}^{\rm BBN+CMB}= 2.880 \pm 0.144 \, .
\label{yofresult}
\eeq
This result is updated in \S\ref{sect:combined} below. 

In this work, we make use of the analyses described in detail in \cite{2001NewA....6..215C,2003PhLB..567..227C,2004PhRvD..70b3505C,Cyburt2016,2020JCAP...03..010F}.
Recently, there has been a burst of activity in response to new
precision measurement of the $d(p,\gamma)\he3$ reaction
by the LUNA collaboration \cite{Mossa2020}.
Recent BBN studies have used two different approaches to nuclear
rates:  an empirical approach, based primarily on experimentally measured cross sections, finds excellent BBN+CMB agreement
\cite{Pisanti2021,2021Yeh},
while an approach incorporating nuclear theory finds some tension \cite{Pitrou2021}.
Additional deuterium reaction measurements are called for to 
resolve this discrepancy \cite{Pitrou2021a}. 
Our work here is an extension of that in ref.~\cite{2021Yeh}
and we note that when the measured nuclear rates are used
there is good agreement with the CMB.  On the other hand,
there is a primordial lithium problem--that is, the predicted BBN \li7/H differs significantly from observations
\cite{2008JCAP...11..012C,Fields2011}. However, much of the evidence in support of associating the \li7 observed in low metallicity halo stars with primordial Li, has evaporated \cite{FO2022}.
Recent non-observations of \li6 in halo stars provide
new evidence that stellar depletion is at play, and thus offer new support for a stellar solution to the problem \cite{FO2022}.
As such, in this paper we 
assume that the Li problem
solution lies not in new physics, but elsewhere,
most likely astrophysical--stellar depletion 
see e.g. \cite{1998ApJ...502..372V,1999ApJ...527..180P,2002ApJ...574..398P,2005ApJ...619..538R,2006Natur.442..657K,2009A&A...503..545L,2016A&A...589A..61G,Deal:2021efl}.

We calculate our results using a series of interlocking codes, beginning with the BBN calculation of light-element abundances, which form the basis for  Monte Carlo runs that inform the likelihood functions described below.
Our BBN code descends from the original Wagoner code \cite{Wagoner1969}, with updates for higher-order corrections to the weak rates \cite{Walker1991} and for integration accuracy \cite{Kernan1993}, 
as described more fully in ref.~\cite{Olive2000}.  The Monte Carlo calculation and likeihood analysis stems from refs.~\cite{2002APh....17...87C,2004PhRvD..70b3505C}, and nuclear reaction rates have been continually updated, most recently in ref.~\cite{2021Yeh} as well as $d+d$ rates as discussed below.  A comparison with work by other groups shows that when the same nuclear rates are adopted, all light-element predictions in are in excellent agreement \cite{GomezInesta2017}.

In this paper we build on the analysis of 
\cite{2020JCAP...03..010F,2021Yeh}. We use updated nuclear rates for $d(d,n)$\he3 and $d(d,p)t$, as well as the updated \he4 abundance \cite{2021JCAP...03..027A} and neutron mean-life \cite{2022RPP}. As in \cite{Cyburt2005,Steigman:2010pa,Mangano:2011ar,2011arXiv1112.2683N}, we concentrate on NBBN
to first set constraints on $N_\nu$ and 
apply these constraints to a wide variety of 
extensions of the Standard Model. 
We focus on models where these extensions can be described as
perturbations away from the SBBN with $N_\nu^{\rm SM}=3$ neutrino species.
In some scenarios there can be changes to the microphysics of element formation due to new particles or interactions and
these require a dedicated analysis beyond the scope of this paper.

In what follows, we perform a combined CMB+NBBN likelihood analysis in order to constrain $N_\nu$ and a host of extensions beyond the Standard Model. As we will see, 
D/H is now important in determining $N_\nu$, 
and as such, we 
first update nuclear rates for deuterium in \S\ref{sect:nukerates}. We
then provide in \S\ref{sect:indepdendent} a brief summary of our  likelihood analysis providing first limits from NBBN separately from those derived from the CMB.  
Next in \S\ref{sect:combined}, we provide our combined (CMB-NBBN) likelihood analysis and  consider traditional limits on $N_\nu$ where no new physics is assumed
between the epoch of BBN and CMB decoupling.
We also allow different values of $N_\nu$ between BBN and CMB decoupling in \S\ref{sect:difference}.
In \S\ref{sect:future} we anticipate the ability of future CMB-S4 and astronomical
\he4 measurements to further sharpen our measures of $N_\nu$, possibly 
even
revealing the predicted standard neutrino heating effects.
Our summary and conclusions are given in \S\ref{sect:conclude}.

\section{Re-evaluation of $d(d,n)\he3$ and $d(d,p)t$ Thermonuclear Rates}
\label{sect:nukerates}

It has long been known that $d(p,\gamma)\he3$, $d(d,n)\he3$, and $d(d,p)t$ dominate the error of the BBN deuterium prediction \cite{Cyburt2016}. Until recently, $d(p,\gamma)\he3$ was the major source of the D/H uncertainty due to its sparse cross section data with relatively large errors at BBN energies \cite{2021Yeh}. However, the precision cross section measurement from the LUNA Collaboration \cite{Mossa2020} revolutionized the evaluation of the $d(p,\gamma)\he3$ thermonuclear rate and its uncertainty. Including the LUNA data in our world average, we reported a new $d(p,\gamma)\he3$ rate with a factor of 2 improvement in its uncertainty, and went on to sow the impact of this new rate on the BBN prediction for D/H as well as constraints on relevant cosmological parameters \cite{2021Yeh}. Similar studies were performed in \cite{Pisanti2021,Pitrou2021}. 

Given the importance of precision D/H calculations for $\eta$ and (to a lesser extent) for $N_\nu$ determinations, we want to evaluate all of the important deuterium destruction rates on the same footing. Prior to the new LUNA data, we had adopted deuterium rates based on NACRE-II \cite{2013NuPhA.918...61X}. Having the updated $d(p,\gamma)\he3$ rate that incorporates the new LUNA data, we will now go one step further; we apply the same methodology and global fitting procedure of ref. \cite{2021Yeh} to re-evaluate $d(d,n)\he3$ and $d(d,p)t$ rates and uncertainties in this study. In this section we summarize the results of this analysis.

\begin{center}
  \begin{figure}[htb]
 \includegraphics[width=0.475\textwidth]{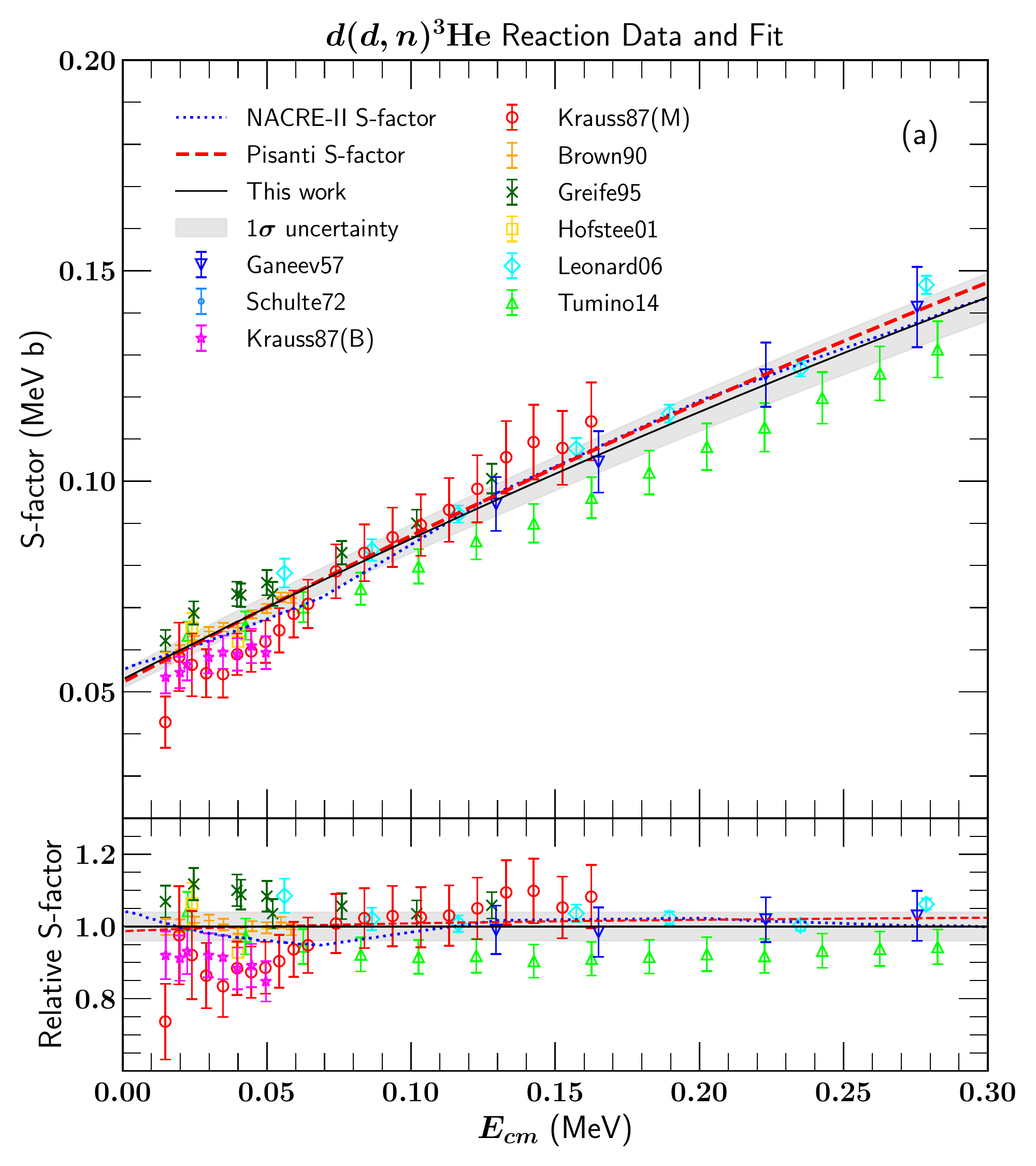}
  \includegraphics[width=0.475\textwidth]{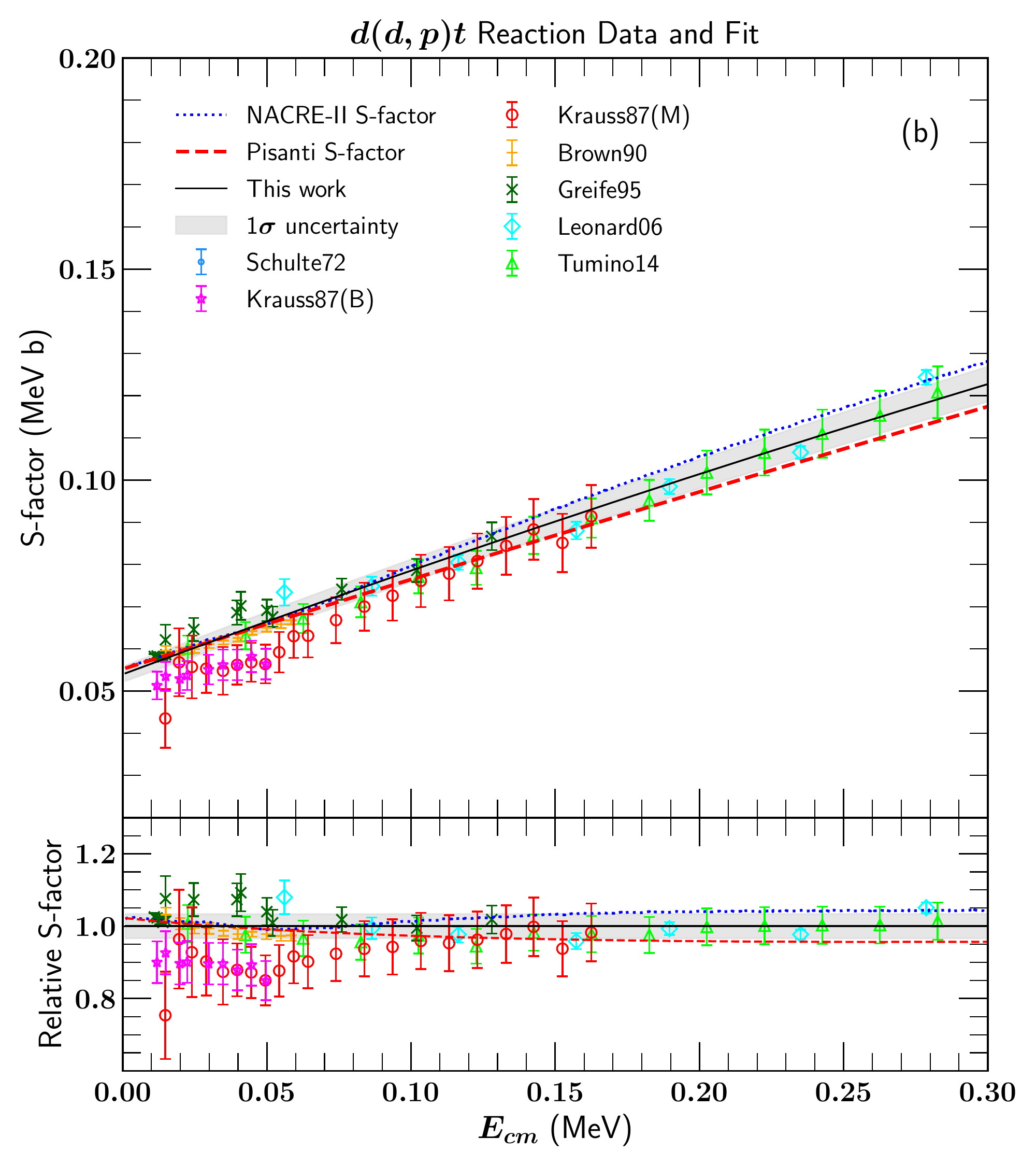}
    \caption{The astrophysical $S$-factor for $d(d,n)$\he3 in the left panel and $d(d,p)t$ in the right panel. On a linear energy scale centered at the BBN energies, we show for both $d+d$ rates 1) the NACRE-II \cite{2013NuPhA.918...61X} $S$-factor used in our previous BBN analysis \cite{2020JCAP...03..010F} (blue dotted); 2) the global average from Pisanti et al. \cite{Pisanti2021} (red dashed); and 3) our new world average rate (black solid).  }
    \label{fig:sfac_linear}
  \end{figure}
\end{center}

\begin{center}
  \begin{figure}[htb]
 \includegraphics[width=0.475\textwidth]{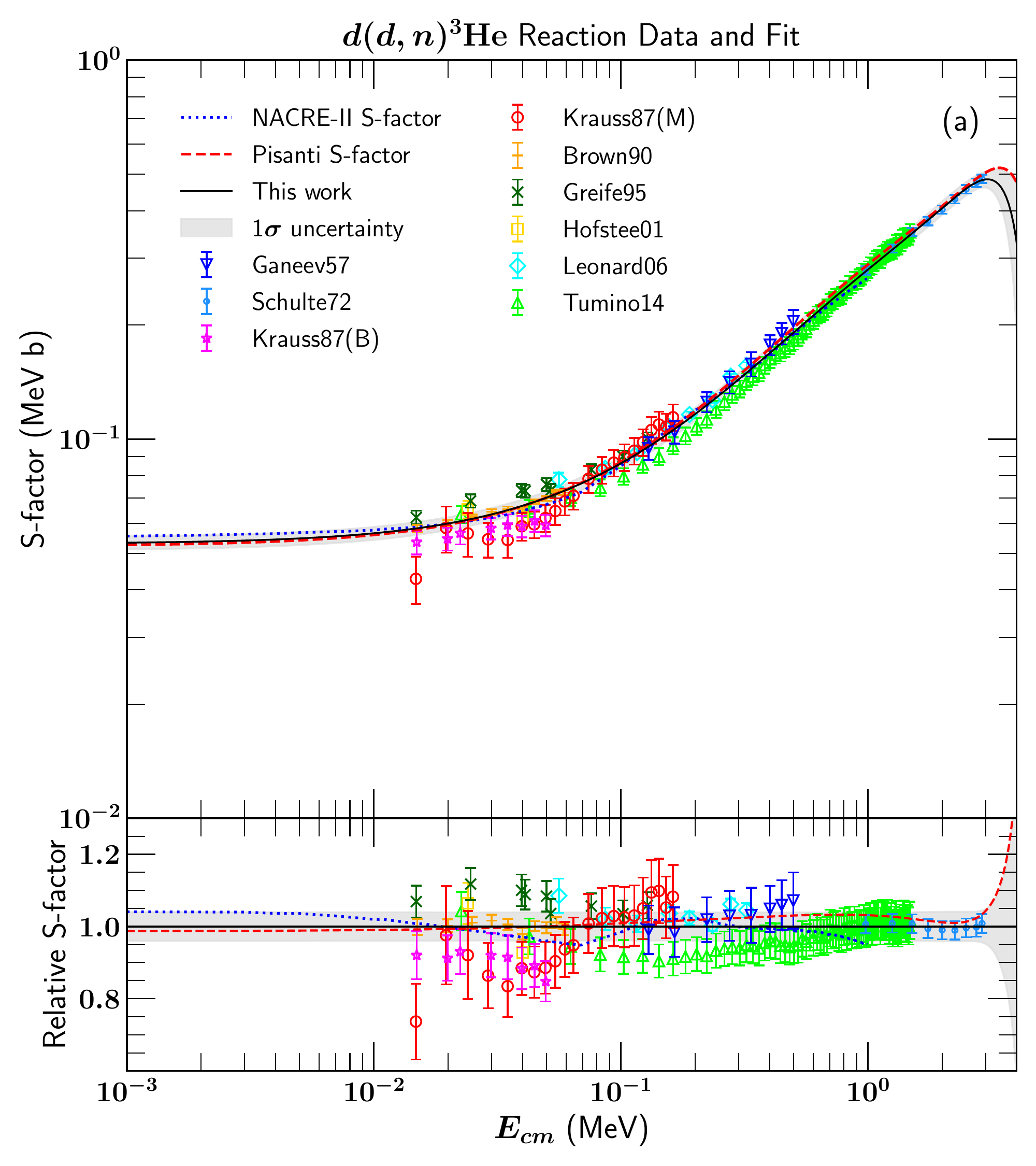}
  \includegraphics[width=0.475\textwidth]{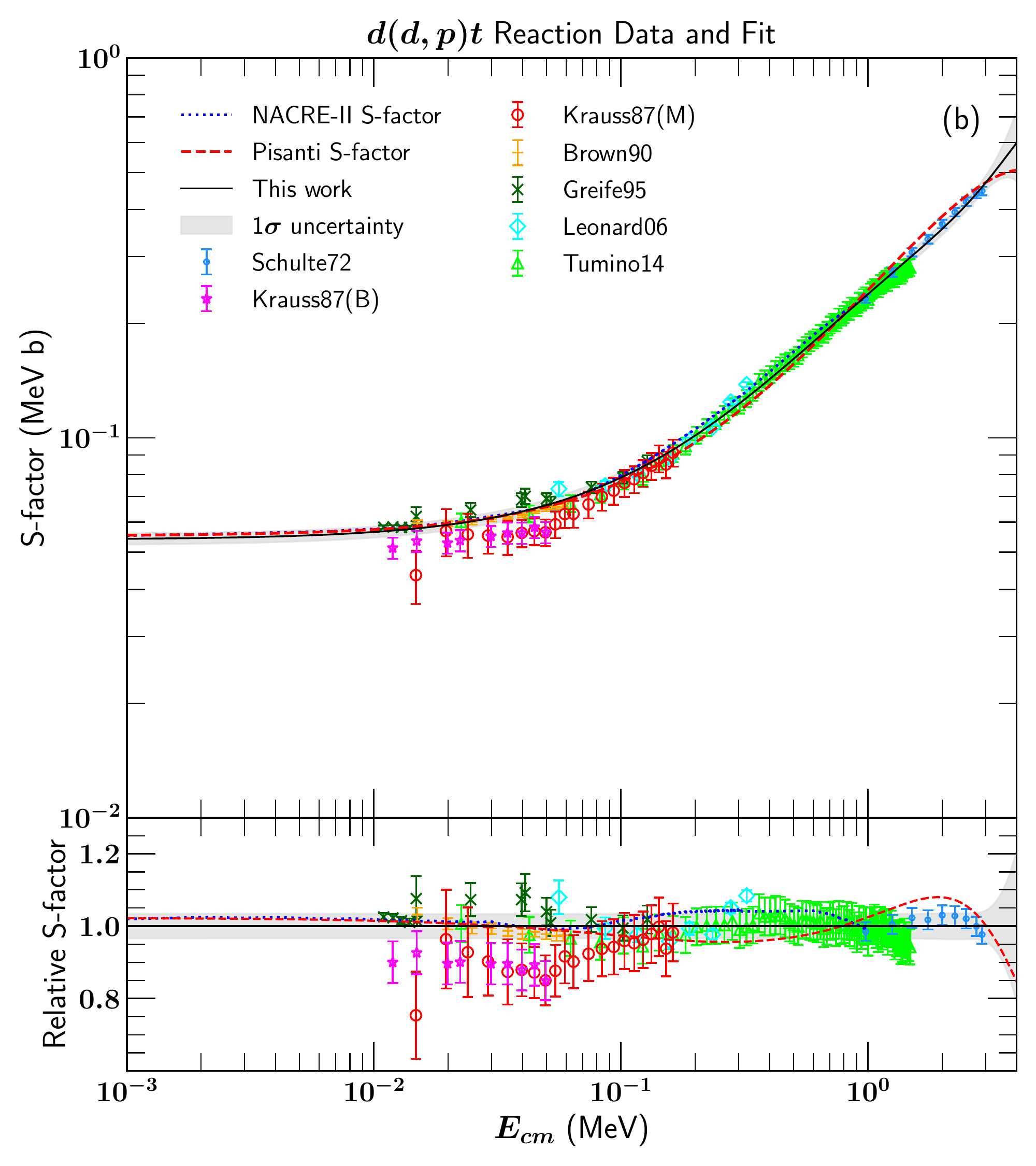}
    \caption{The astrophysical $S$-factor for $d(d,n)$\he3 in the left panel and $d(d,p)t$ in the right panel on a log energy scale centered at the BBN energies. }
    \label{fig:sfac_log}
  \end{figure}
\end{center}

Fortuitously, $d(d,n)\he3$ and $d(d,p)t$ share the same initial state,
and so can be measured in the same experiment. Because we aim to re-evaluate these two $d+d$ rates with our new procedure, we use the similar datasets adopted in NACRE-II \cite{2013NuPhA.918...61X}.
We include cross section data for $d(d,n)\he3$ and $d(d,p)t$ from
Schulte et al. \cite{Schulte:1972enh}, Krauss et al. \cite{Krauss:1987nvu}, Brown et al. \cite{PhysRevC.41.1391}, Greife et al. \cite{1995ZPhyA.351..107G}, Leonard et al. \cite{Leonard:2006cd}, and Tumino et al. \cite{Tumino:2011zz}.  In addition, we add data from Ganeev et al. \cite{Ganeev1958} and Hofstee et al. \cite{2001NuPhA.688..527H}, 
who only report results for  $d(d,n)\he3$.
To avoid the impact of the laboratory electron screening effect at low energies, we adopt data points above 10 keV, which is also below the Gamow window for $d+d$.\footnote{At $T_9=1$ ({\it i.e.,} $T = 10^9$ K), the asymmetric Gamow peak for $d+d$ has maximum value at 122 keV and its window at 1/e height is within (40, 300) keV.} For the high energy end, we stop at 3 MeV because it sufficiently covers the most important energy ranges for both BBN $d+d$ rates. These adopted data are plotted in Figures \ref{fig:sfac_linear} and \ref{fig:sfac_log} in terms of the astrophysical $S$-factor $S(E) = E\sigma(E)e^{2\pi\eta_s}$ versus the center-of-mass energy $E$ to factor out the effect of the Coulomb barrier, where $\eta_s = e^2/\hbar v$ is the Sommerfeld parameter and $v = v(E)$ is the relative velocity of reactant D nuclides, and $\sigma(E)$ is the reaction cross section.
In Figure \ref{fig:sfac_log},  we show the $S$-factor for the $d+d$ reactions as in Fig.~\ref{fig:sfac_linear}, but for wider and logarithmic energy scale.
We see that our fits agrees with the experimental data out to $\sim 2 \ \rm MeV$.  Larger energies are sufficiently beyond the range of validity of our 4th-order polynomial description that the fit becomes unphysical.

Following the same nuclear cross section fitting procedure developed for $d(p,\gamma)\he3$ \cite{2021Yeh}, we fit the $S$-factor using a series of polynomials in terms of center-of-mass energy $E$. Because both $d(d,n)\he3$ and $d(d,p)t$ $S$-factor plots show smooth behavior around BBN energy range, we found that a polynomial expansion including a 4th-order ($E^4$) term agrees well with the data without overfitting for both reactions. The global best fit is determined by $\chi^2$ minimization. Beyond the experimental energy-dependent uncertainty (statistical and systematic errors combined in quadrature), we also include an energy-independent uncertainty to account for the systematic discrepancies among datasets \cite{2001NewA....6..215C}. The resulting $\chi^2$ per degree of freedom is around unity with the inclusion of such a discrepancy error.

We also present in Figures \ref{fig:sfac_linear} and \ref{fig:sfac_log} relevant $S$-factor fits for both $d(d,n)\he3$ (on the left panel) and $d(d,p)t$ (on the right panel) considered in this paper as a function of energy. The blue dotted curve is from NACRE-II \cite{2013NuPhA.918...61X} and was adopted in our previous BBN studies \cite{Cyburt2016,2020JCAP...03..010F,2021Yeh}. The black solid curve is our new average rate with the grey shaded region for our calculated 1-$\sigma$ uncertainty. For comparison, the work done here is basically re-evaluating the $d+d$ rates based on a dataset selection similar to NACRE-II but using our own fitting procedure. Figure \ref{fig:sfac_linear} shows that the NACRE-II and our new cross section fits agree with each other and the data within uncertainties at BBN energies. Moreover, Pisanti et al. have recently reported their latest fits for both $d+d$ rates based on a similar empirical polynomial fitting procedure \cite{Pisanti2021}. We include their fit in Figures \ref{fig:sfac_linear} and \ref{fig:sfac_log} using the red dashed curve for data analysis method comparison. In the lower portion of these two panels, the data and other fits are shown relative to new fit used here.

Once we have the best fit for the $S$-factor, we can calculate the average thermonuclear rate as a function of temperature $T$ using
\begin{equation}
    \lambda(T) = N_{\rm Avo} \langle \sigma v \rangle = N_{\rm Avo} \bigg( \frac{8}{\mu \pi}\bigg) ^{1/2} (kT)^{-3/2}\int^{E_{\rm max}}_{E_{\rm min}}S(E)e^{-2\pi \eta_{s}} e^{-E/kT} dE,
    \end{equation}
where $N_{\rm Avo}$ is Avogadro's number and $\mu$ is the reduced mass. The integration bounds are set to be ($E_{\rm min}$, $E_{\rm max}$) = (0 $kT$, 100 $kT$) for practical calculation.

\begin{center}
  \begin{figure}[htb]
  \centering
 \includegraphics[width=0.8\textwidth]{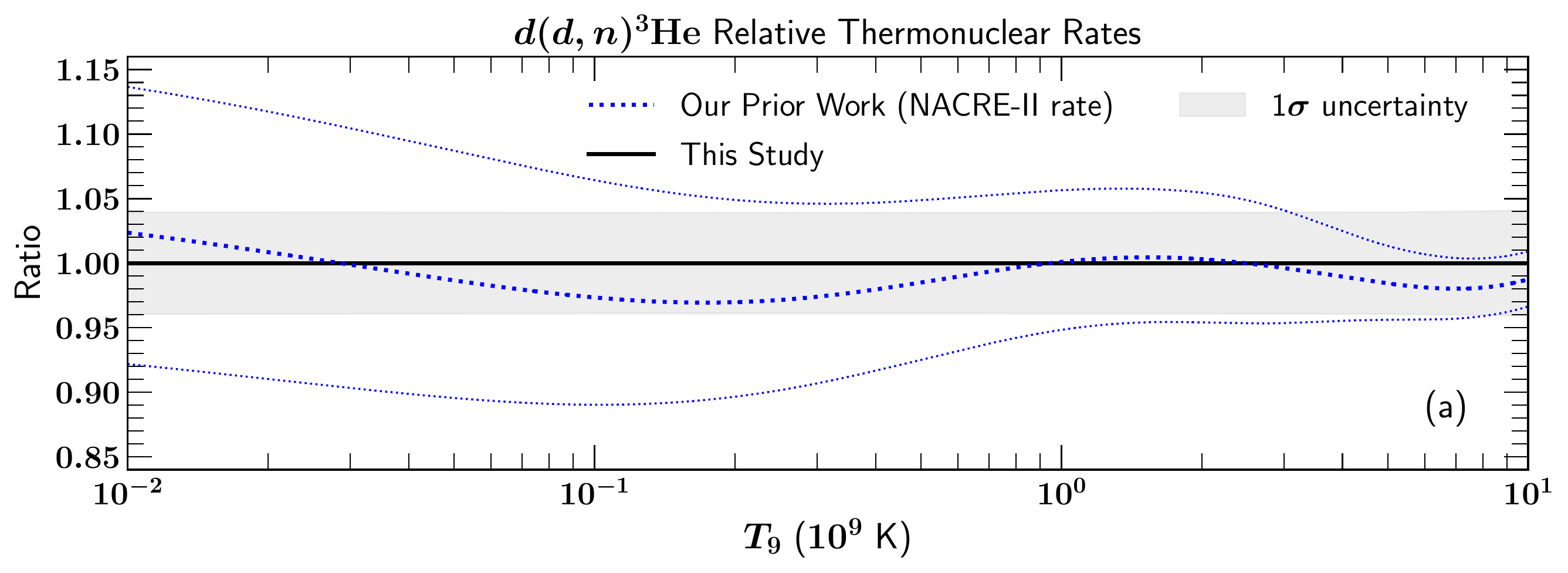}\\
  \includegraphics[width=0.8\textwidth]{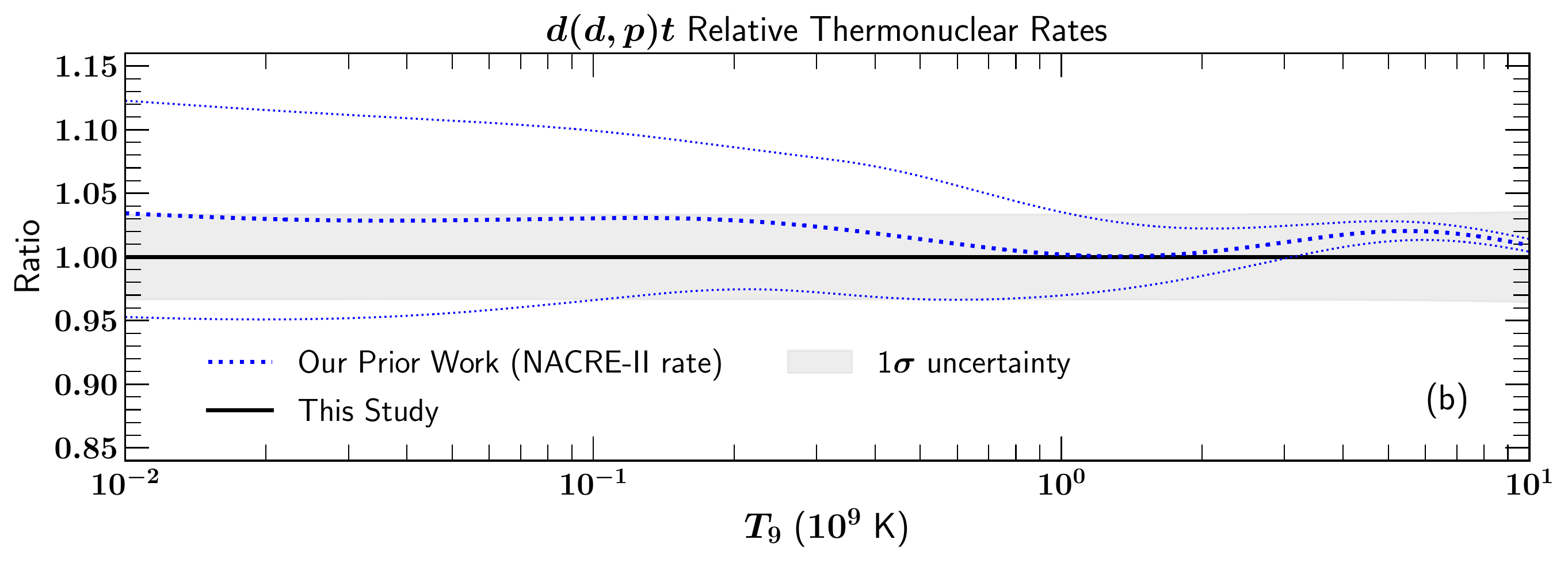}
    \caption{The comparison between the old thermonuclear rates adopted in our prior work \cite{Cyburt2016,2020JCAP...03..010F,2021Yeh} and our new averages as baselines for $d(d,n)$\he3 in the left panel and $d(d,p)t$ in the right panel. The old rates, calculated from NACRE-II \cite{2013NuPhA.918...61X}, are in dotted blue with 1$\sigma$ bounds. The 1$\sigma$ uncertainties of new rates are indicated by the grey regions. Curves are shown as a function of temperature in units of $T_9 = 10^9$ K. Our new rates are consistent with the old rates around $T_9 = 1$.}
    \label{fig:relative_rates}
  \end{figure}
\end{center}

In Figure \ref{fig:relative_rates}, using our new rates as baselines, we show the relative $d+d$ thermal rates for our previous work \cite{Cyburt2016,2020JCAP...03..010F,2021Yeh} calculated from the fits of NACRE-II. Our new rates are consistent with NACRE-II rates around $T_9 = 1$, which is at the heart of BBN deuterium synthesis. In fact, within uncertainty, the $d(d,n)\he3$ rate from NACRE-II is a few percent smaller than our new baseline at $0.1 < T_9 < 1$, but for $d(d,p)t$ the trend goes in the opposite direction. Because these two $d+d$ rates have similar scaling relations to the deuterium prediction \cite{2021Yeh}, the downward and upward differences shown in the panels of Figure \ref{fig:relative_rates} at $0.1 < T_9 < 1$ almost cancel in practice. Thus we do not expect large changes in the mean values of D/H calculated with our updated network versus the our older work. But in our new analysis, the predicted deuterium uncertainty contributions have decreased by a factor of $\sim1.5$ from $d(d,n)\he3$ and a factor of $\sim1.4$ from $d(d,p)t$. Therefore, we anticipate that the the resulting full D/H uncertainty will be lower by a similar factor. 

\subsection{Primordial Light Element Abundance Predictions}
Using the baryon density from the \verb+base_yhe_plikHM_TTTEEE_lowl_lowE_post_lensing+ chains of {\em Planck 2018} data as input for the $N_\nu=3$ standard case,\footnote{These are CMB MCMC chains based on TT, TE and EE (temperature and polarization) power spectra, including lensing reconstruction and low $\ell$ multipoles for analyses.} here we list our latest SBBN light element abundance predictions with their means and $1\sigma$ errors:\footnote{We also include now an updated neutron mean lifetime which is $\tau_n = 878.4 \pm 0.5$s \cite{UCNt:2021pcg,2022RPP}.} 
\begin{eqnarray}
Y_p &=& 0.2467\pm0.0002,\\
{\rm D/H} &=& (2.506\pm0.083)\times 10^{-5},\\
\he{3}/{\rm H} &=& (10.45\pm 0.87)\times 10^{-6},\\
\li{7}/{\rm H} &=& (4.96\pm 0.70)\times 10^{-10}.
\end{eqnarray}
We remind readers that 
these (and Table \ref{tab:eta} below) are the only results in this paper where we used the standard CMB (fixed $N_{\rm eff}$) chains. Compared with Equation (\ref{yofdh}), our new predicted ${\rm D/H}$ uncertainty has been improved by a factor of $\sim 1.3$ relative to \cite{2021Yeh}, as expected.\footnote{In both studies, we used the same {\em Planck 2018} MCMC chain and the same thermonuclear rates except $d(d,n)\he3$ and $d(d,p)t$.} However, it still falls behind the observational uncertainty shown in Equation (\ref{dhobs}). Currently, $d(d,n)\he3$ dominates the deuterium error budget in our study by contributing $\sigma(D/H) = 0.053\times 10^{-5}$, followed by $d(d,p)t$ with $\sigma(D/H) = 0.039\times 10^{-5}$ and then $d(p,\gamma)\he3$  with $\sigma(D/H) = 0.036\times 10^{-5}$.\footnote{These uncertainties are evaluated at a fixed $\eta = 6.104\times10^{-10}$ determined from the {\em Planck} chain mentioned above.} To further improve the BBN deuterium calculation after LUNA's precision measurements of $d(p,\gamma)\he3$, future precision cross section measurements for $d(d,n)\he3$ and $d(d,p)t$ at BBN energies are now desired.

For completeness, we show in Table \ref{tab:eta}, updated results for our determination of $\eta$ from the CMB alone and in combination with BBN using various combinations of light element observations when $N_\nu = 3$ is fixed. The likelihood functions used to obtain these results were defined in \cite{2020JCAP...03..010F} and can be inferred from the likelihood functions defined below. 

\begin{table}[!htb]
\caption{Constraints on the baryon-to-photon ratio, using different combinations of observational constraints.  We have marginalized over $Y_p$ to create 1D $\eta$ likelihood distributions with $N_\nu = 3$. Given are both the mean (and its uncertainty)
as well as the value of $\eta$ at the peak of the distribution.
\label{tab:eta}
}
\vskip .2in
\begin{center}
\begin{tabular}{|l|c|c|}
\hline
 Constraints Used & mean $10^{10} \eta$ & peak $10^{10} \eta$ \\
\hline
CMB-only & $6.104\pm 0.055$ & 6.104 \\
\hline
\hline
BBN+$Y_p$ & $6.239 {}^{+1.202}_{-2.741}$ & 5.031 \\
\hline
BBN+D & $6.042\pm 0.118$ & 6.041 \\
\hline
BBN+$Y_p$+D & $6.040\pm 0.118$ & 6.039 \\
\hline
CMB+BBN & $6.124\pm 0.040$ & 6.124 \\
\hline
CMB+BBN+$Y_p$ & $6.124\pm 0.040$ & 6.124 \\
\hline
CMB+BBN+D & $6.115\pm 0.038$ & 6.115 \\
\hline
\hline
CMB+BBN+$Y_p$+D & $6.115\pm 0.038$ & 6.115 \\
\hline
\end{tabular}
\end{center}
\end{table}

We use the {\em Planck} \verb+base_nnu_yhe_plikHM_TTTEEE_lowl_lowE_post_lensing+ chains for NBBN abundances; these include temperature and polarization data, as well as lensing. For the completeness of this section, here are the abundance predictions when $N_\nu$ is not fixed:
\begin{eqnarray}
Y_p &=& 0.2439\pm0.0041,\\
{\rm D/H} &=& (2.447\pm0.117)\times 10^{-5},\\
\he{3}/{\rm H} &=& (10.37\pm 0.88)\times 10^{-6},\\
\li{7}/{\rm H} &=& (5.03\pm 0.72)\times 10^{-10}.
\end{eqnarray}

\section{Independent Limits on $N_\nu$ from the BBN and the CMB}
\label{sect:indepdendent}

In this section we present independent BBN and CMB limits on $\eta$ and $N_\nu$, using 
likelihood analyses.  
This allows us to compare these measures, which provides a first assessment of the 
consistency between these results.  Agreement would support the standard scenario where
both $\eta$ and $N_\nu$ are unchanged after BBN.  Discrepancy could point to new physics
between the BBN and CMB epochs.

We compute the BBN likelihood functions following the formalism we have described
elsewhere, e.g., \cite{Cyburt2016,2020JCAP...03..010F}; here we summarize the key results.
BBN theory as embodied in our code predicts light element abundances $\vec{X} = (Y_p, {\rm D/H, \he3/H}, \li7/{\rm H})$ for each choice of the pair $(\eta,N_\nu)$ and nuclear reaction rates.
Varying nuclear reaction rates within their uncertainties via a Monte Carlo 
gives the likelihood function ${\cal L}_{\rm NBBN}(\vec{X} ; \eta,N_\nu)$. 

Astronomical observations determine the 
abundances $X_i$ for each light nuclide $i$, giving likelihoods ${\cal L}_{{\rm obs}}(X_i)$ which we model as Gaussians.
We convolve with the BBN predictions to infer
\begin{equation}
\label{eq:LBBN}
{\cal L}_{\rm NBBN+obs}(\eta, N_\nu)  \ \propto \ \int  {\cal L}_{\rm NBBN}(\vec{X} ; \eta,N_\nu) \ \prod_i {\cal L}_{{\rm obs}}(X_i)  \ dX_i \ \ .
\end{equation}
We implicitly assume flat priors for $\eta$ and $N_\nu$.

As noted in our previous work, not
all light element abundances are in practice suitable as probes of cosmology.
\he3 does not have a sufficiently well-measured primordial abundance
(\cite{Olive1995,Scully:1995tp,Olive:1996tt,Vangioni2003}, but see \cite{Cooke:2022cvb}),
and there are multiple reasons to suspect that  \li7 observations do not
reflect the primordial abundance \cite{FO2022}.
Therefore the abundance observations available for
our analysis are ${\rm D/H}$ and $Y_p$, so
the product in Eq.~(\ref{eq:LBBN}) 
can have one or two terms depending on which of these one uses.

Turning to the CMB constraints, we use the likelihoods derived from the final {\em Planck} 2018 analysis. 
These likelihoods depend on $\eta$, $N_{\rm eff}$ and $Y_p$.\footnote{A slightly different convention for the definition of the helium mass fraction, $Y_p$, is adopted in the {\it Planck} MCMC chains. We convert it to the BBN $Y_p$ convention using Appendix A of ref. \cite{2020JCAP...03..010F}.}  This likelihood is sensitive to the primordial (elemental)
helium abundance, because the damping tail is sensitive
to the number of electrons per baryon,
which in turn depends on $Y_p$.
To preserve independence from BBN, we use the MCMC chains that do not use the nucleosynthesis
relationship giving $Y_p$ at each $\Omega_{\rm b}h^2$.
This likelihood is well fit by
correlated Gaussians with small high-order corrections \cite{Cyburt2016}.
As mentioned in \S\ref{intro}, the CMB measures $N_{\rm eff}$, which slightly differs from the BBN value due to 
heating effects during BBN at neutrino freeze-out.  For the standard $N_\nu=3$ case this gives
$N_{\rm eff} = 3.044$ \cite{Akita:2020szl,Bennett:2020zkv,EscuderoAbenza:2020cmq,Froustey:2020mcq},
which we extend to 
\begin{equation}\label{NeffNnu}
    N_{\rm eff} = (1.0147)N_{\nu}.
\end{equation}
We thus arrive at a CMB likelihood ${\cal L}_{\rm NCMB}(\eta,N_\nu,Y_p)$.
By marginalizing the CMB likelihood over $Y_p$, we can obtain
\begin{equation}
{\mathcal L}_{\rm NCMB}(\eta, N_\nu) \ \propto \  \int  {\mathcal L}_{\rm NCMB}(\eta,N_\nu,Y_p) \ dY_p \, .
\label{LNCMB}
\end{equation}

\begin{figure}[!ht]
    \centering
    \includegraphics[width=\textwidth]{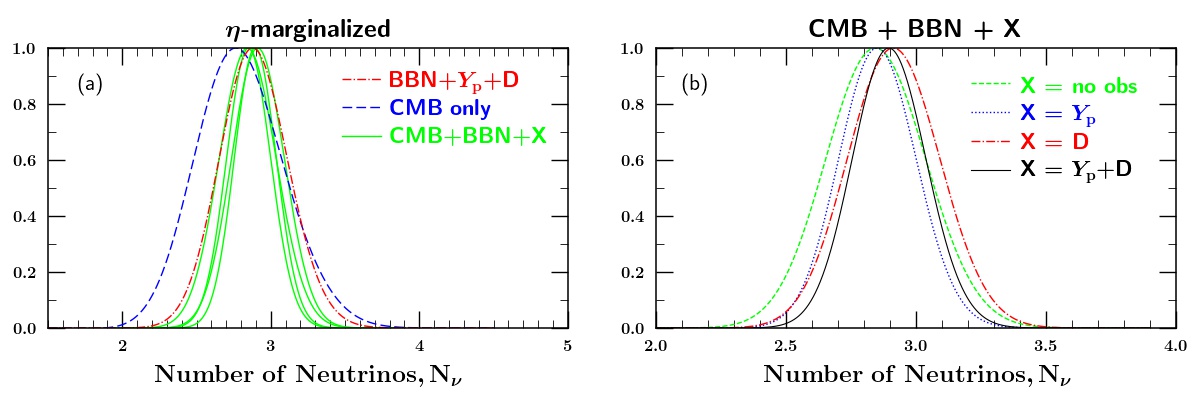}
    \caption{Likelihood distributions for $N_\nu$ for
    BBN and the CMB separately and combined.
    In all cases shown, the likelihood function has been marginalized over the baryon-to-photon ratio $\eta$.
    (a) BBN-only, CMB-only, and combined limits.  (b) Zoom into BBN+CMB joint limits to highlight results for different combinations of light element abundances. The last three columns of
    Table \ref{tab:etannu} summarize these results.}
    \label{fig:N_nu_dist_Aver}
\end{figure}

Figure \ref{fig:N_nu_dist_Aver} and
Table \ref{tab:etannu} present limits on $N_\nu$
for BBN and the CMB independently, and combined under
the assumption that relevant cosmological parameters are the same at the two epochs (which is addressed further in \S \ref{sect:combined}).
The BBN-only limits marginalize over 
$\eta$ and light element observations D/H and/or $Y_p$
\begin{equation}
{\cal L}_{\rm NBBN+obs}(N_\nu) \ \propto \ \int {\cal L}_{\rm NBBN} (\vec{X}; \eta,N_\nu) \  d\eta \  \prod_{i} {\cal L}_{\rm obs}(X_i) \ d X_i \, .
\end{equation}
Here and below, the product over observations
includes one or two terms depending on the D/H and $Y_p$ combination used.
The result appear as the dot-dashed red curve in the left panel of Figure~\ref{fig:N_nu_dist_Aver}, where we see it peaked near, but slightly below, the Standard Model value.  The peak value and mean value of $N_\nu$ for this case are given in the 2nd row of Table
\ref{tab:etannu}.

\begin{table}[!htb]
\caption{The separately marginalized central 68.3\% confidence limits and most-likely values on the baryon-to-photon ratio $\eta$ and effective number of neutrinos $N_\nu$, using different combinations of observational constraints. The 95.45\% upper limits from Eq.~(\ref{int}), given that $ N_{\nu}>3$, are also shown in the last column.
\label{tab:etannu}
}
\vskip.1in
\begin{center}
\begin{tabular}{|l||c|c||c|c||c|}
\hline
 Constraints Used & mean $\eta_{10}$ & peak $\eta_{10}$ & mean $N_\nu$ & peak $N_\nu$ & $\delta N_{\nu}$ \\
\hline\hline
CMB-only & $6.090 \pm 0.061$ &  $6.090^{   +0.061}_{ -0.062}$ & $2.800 \pm 0.294$ &  $2.764^{   +0.308}_{ -0.282}$ & 0.513\\
\hline
BBN+$Y_p$+D & $5.986 \pm 0.161$ &  $5.980^{   +0.163}_{ -0.159}$ & $2.889 \pm 0.229$ &  $2.878^{   +0.232}_{ -0.226}$ & 0.407\\
\hline
CMB+BBN & $6.087 \pm 0.061$ &  $6.088^{   +0.061}_{ -0.062}$ & $2.848 \pm 0.190$ &  $2.843^{   +0.192}_{ -0.189}$ & 0.296\\
\hline
CMB+BBN+$Y_p$ & $6.089 \pm 0.053$ &  $6.089^{   +0.054}_{ -0.054}$ & $2.853 \pm 0.148$ &  $2.850^{   +0.149}_{ -0.148}$ & 0.221\\
\hline
CMB+BBN+D & $6.092 \pm 0.060$ &  $6.093^{   +0.061}_{ -0.060}$ & $2.916 \pm 0.176$ &  $2.912^{   +0.178}_{ -0.175}$ & 0.303\\
\hline
CMB+BBN+$Y_p$+D & $6.088 \pm 0.054$ &  $6.088^{   +0.054}_{ -0.054}$ & $2.898 \pm 0.141$ &  $2.895^{   +0.142}_{ -0.141}$ & 0.226\\
\hline
\end{tabular}
\end{center}
\end{table}

It is also possible to marginalize over $N_\nu$ to obtain a likelihood as a function of $\eta$, 
\begin{equation}
{\cal L}_{\rm NBBN+obs}(\eta) \ \propto \ \int {\cal L}_{\rm NBBN} (\vec{X}; \eta,N_\nu) \  dN_\nu \  \prod_{i} {\cal L}_{\rm obs}(X_i) \ d X_i \, .
\end{equation}
The mean and peak values of $\eta$ from the BBN-only likelihood function is also given in the 2nd row of Table \ref{tab:etannu}.

For the CMB-only results 
we marginalize the likelihood given in Eq.~(\ref{LNCMB}) over $\eta$
to obtain the distribution in $N_\nu$
\begin{equation}
{\mathcal L}_{\rm NCMB}(N_\nu) \ \propto \  \int  {\mathcal L}_{\rm NCMB}(\eta,N_\nu) \ d\eta \, .
\end{equation}
This appears as the dashed blue curve in the left panel of Figure \ref{fig:N_nu_dist_Aver}, which is
entirely consistent with the BBN-only curve and the Standard Model value, though the peak lies slightly below both. The mean and peak values of $N_\nu$ from the CMB-only likelihood function is given in the first row of Table~\ref{tab:etannu}. Similarly, we can marginalize over $N_\nu$ to obtain
\begin{equation}
{\mathcal L}_{\rm NCMB}(\eta) \ \propto \  \int  {\mathcal L}_{\rm NCMB}(\eta,N_\nu) \ dN_\nu \, .
\end{equation}
The mean and peak values of $\eta$ from the CMB-only likelihood function is also given in the first row of Table \ref{tab:etannu}.

Figure \ref{fig:N_nu_dist_Aver} shows that the 
BBN and CMB determinations of $N_\nu$ are in excellent 
agreement with each other, and with the Standard Model value.
These three measures are all independent, so their
concordance is by no means guaranteed, but to the
contrary marks a great success of hot big bang cosmology. 
Put differently, this agreement tells us that
{\em BBN and the CMB are consistent with both
the Standard Model
($N_\nu=3$) and standard Cosmology ($N_\nu^{\rm BBN} = N_{\nu}^{\rm CMB}$), showing no need for new physics
within our ability to measure.}

It is also remarkable that BBN and the CMB probes 
$N_\nu$ with similar precision.
The BBN limits remain slightly tighter, 
but the improvement of the CMB constraints 
after {\em Planck} has made them closely
competitive.  The comparable strength in these two measure
now offers new ways to probe the early universe, as we now see.

The agreement between the BBN and CMB measures of $N_\nu$
invites us to press onward in two ways.
(1) We can {\em combine} the BBN and CMB limits on $N_\nu$, assuming nothing occurs between the two epochs to change this parameter. This analysis appears in \S \ref{sect:combined}, and this approach is the one adopted in work to date.   (2)  We now can also search for,
and place limits on,
possible {\em differences} between $N_\nu$ at the BBN and CMB epochs.  This approach is novel, and appears in \S \ref{sect:difference}.

\section{BBN and the CMB Combined: No New Physics After Nucleosynthesis}
\label{sect:combined}

In this section we combine the BBN and CMB constraints
on $N_\nu$, and apply the resulting limits to a variety of
particle physics and astrophysics examples. 
The limits we derive here rest on the assumption that
$N_\nu^{\rm CMB} = N_\nu^{\rm BBN}$, that is, there
is no change in the cosmic radiation content between the
two epochs.
This assumption is relaxed in \S \ref{sect:difference}.
This approach is similar to that used in prior work. Thus our
results here for the best fits to $\eta$ and $N_\nu$ 
are an update of our findings
in \cite{2021Yeh}, with the only  difference
being the updates to the $d+d$ reaction rates as explained in \S\ref{sect:nukerates}, as well as an updated primordial helium mass fraction and neutron mean-life.

\subsection{BBN+CMB Limits on $N_\nu$}

Two dimensional
joint limits on $(\eta,N_\nu)$ are obtained using Eq.~(\ref{eq:LBBN}) for the BBN-only likelihood and Eq.~(\ref{LNCMB}) for the CMB-only likelihood function. 
We can then combine BBN and the CMB to get tighter joint limits on 
both $\eta$ and $N_\nu$.
This was first done in \cite{Cyburt2016} 
and is an extension of the traditional BBN-only approach, e.g., in \cite{Cyburt2005}.
The combined likelihood is
\begin{equation}
    {\cal L}_{\rm NBBN+NCMB+obs}(\eta, N_\nu)
    \ \propto \ 
    \int {\cal L}_{\rm NCMB}(\eta,N_\nu,Y_p) \ 
    {\cal L}_{\rm NBBN}(\vec{X};\eta, N_\nu) \ 
    \prod  {\cal L}_{\rm obs}(X_i) \ dX_i \ .
    \label{BBNCMBobs}
\end{equation}
In cases where \he4 observations are used,
the $Y_p$ marginalization links the BBN
and CMB distribution which are both
sensitive to this value.

\begin{figure}
    \centering
    \includegraphics[width=\textwidth]{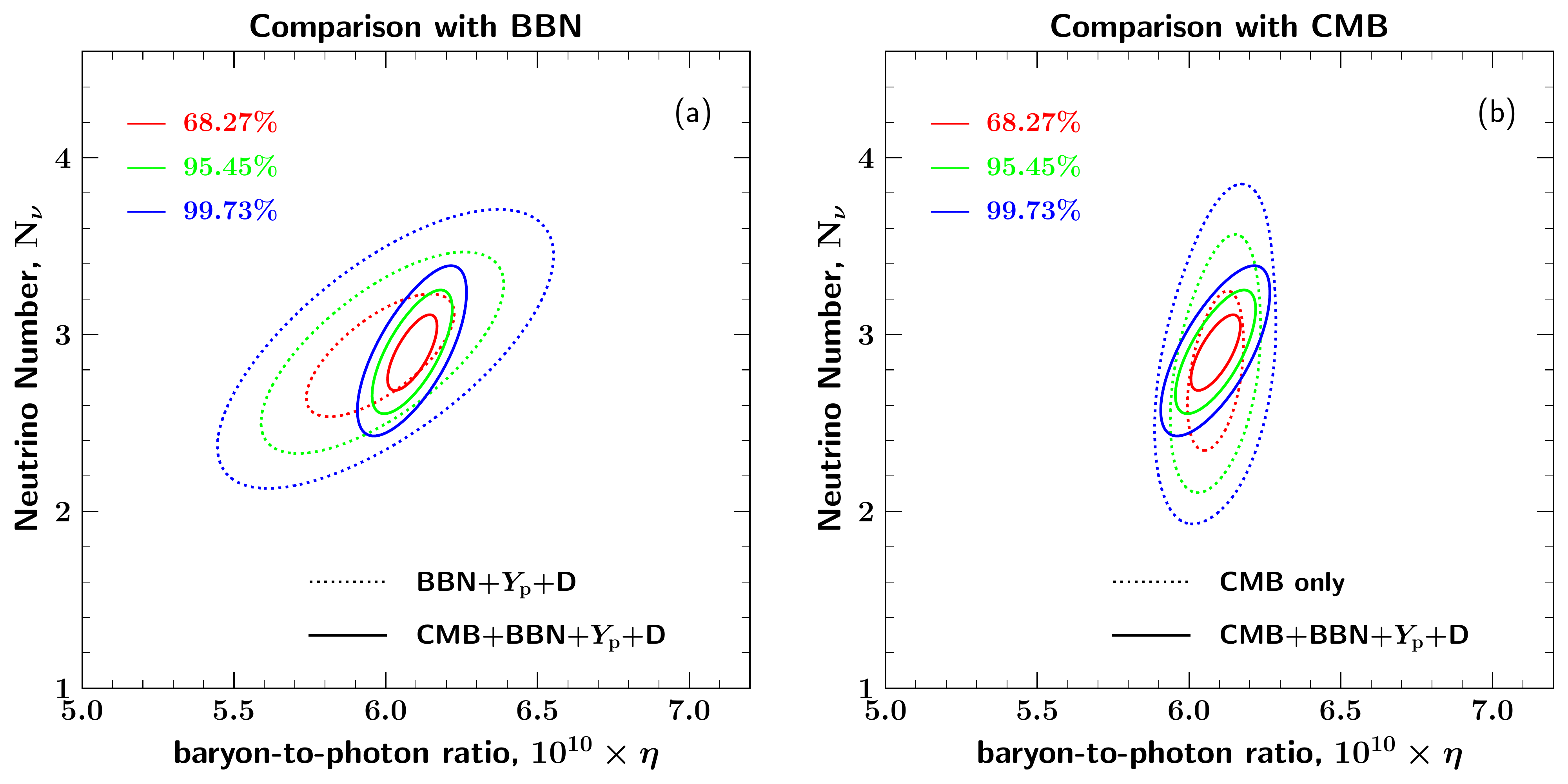}
    \caption{The joint or 2D likelihood ${\cal L}(\eta,N_\nu)$, assuming these parameters do not change between after BBN.  (a) Dashed contours show BBN-only results, while
    solid contours give combined BBN+CMB results. 
    (b) Contours for the CMB only, and again BBN+CMB combined. 
    Overall, we see that the combined BBN+CMB+obs likelihood combines the virtues of the tight CMB limits on $\eta$ and the stronger BBN limits on $N_\nu$.}
    \label{fig:Nnu_vs_eta}
\end{figure}

A projection of the likelihood function (\ref{BBNCMBobs}) onto the $(\eta, N_\nu)$ plane is shown by the solid contours in Fig.~\ref{fig:Nnu_vs_eta}. These are compared with the BBN-only results using Eq.~(\ref{eq:LBBN}) and the CMB-only results using Eq.~(\ref{LNCMB}) in the left and right panels depicted by the dotted contours. 
We see that the BBN-only constraint shows a significant
positive correlation between $\eta$
and $N_\nu$. This arises 
because
${\rm D/H} \propto \eta^{-1.6} N_\nu^{0.4}$
and so for fixed $({\rm D/H})_{\rm obs}$,
we see the positive correlation 
$N_{\nu} \propto \eta^{4}$
follows.
On the other hand, for the CMB
there is very little correlation between
$\eta$ and $N_\nu$.
Both the BBN-only and CMB-only results are consistent with each other,
justifying their combined use.
Figure~\ref{fig:Nnu_vs_eta} displays that BBN provides a slightly better $N_\nu$
determination, while the CMB dominates the measurement of $\eta$.
This illustrates a BBN-CMB complementarity.
As one can see, results are also in excellent agreement with the Standard Model value of $N_\nu=3$.

If we marginalize (\ref{BBNCMBobs}) over $\eta$, we obtain the 
distributions over $N_\nu$ which
appear in Figure~\ref{fig:N_nu_dist_Aver}
for different light element observation
combinations. In
panel (a) these appear as the solid curves which are zoomed in on in 
panel (b) which
shows the effect of combining the CMB with
different BBN+obs choices; this 
moderately improves on the BBN-only 
constraints.
As one would expect,
progressively 
adding light element observations 
leads to further modest improvements on
the $N_\nu$ limit.

Note that the case labeled `$X =$ no obs' 
corresponds to $\int {\cal L}_{\rm CMB}(\eta,N_\nu,Y_p) \ {\cal L}_{BBN}(Y_p; \eta,N_\nu) \ d\eta \ dY_p$
where we use no astronomical observational data for $Y_p$ (or D/H),
but through ${\cal L}_{\rm BBN}$ we
introduce the BBN theory connection
among $Y_p$, $\eta$, and $N_\nu$.
This case gives a noticeably stronger
limit on $N_\nu$ than
those of the CMB only.  
This demonstrates another aspect 
of BBN-CMB complementarity,
as the tight BBN predictions for $Y_p$ 
alleviate the weaker CMB sensitivity to
this parameter.

From Fig.~\ref{fig:N_nu_dist_Aver} and Table~ \ref{tab:etannu}
we see that $Y_p$ has the stronger impact on the BBN+CMB constraints,
but that D/H still has an impact despite playing a lesser role.
Improved astronomical $Y_p$ measurements, while quite challenging,
would thus improve $N_\nu$ limits, as we will
see below in detail (\S \ref{sect:future}).  

Our best limit uses both light elements and the CMB:
\begin{equation}
\label{eq:Nnu-tout}
    N_\nu = 2.898 \pm 0.141
\end{equation}
i.e, the case CMB+BBN+$Y_p$+D/H.
This gives a 2$\sigma$ upper limit of
\begin{equation}
\label{eq:DeltaNnu2side}
    \Delta N_\nu < 0.180
\end{equation}
arising from our two-sided error range about the mean in Eq.~(\ref{eq:Nnu-tout}).
This result updates that given in Eq.~(\ref{yofresult})
by the inclusion of the newly adopted deuterium rates, primordial helium abundance, and neutron mean-life.
As one can see, these updates are incremental hopefully indicating the robustness of the limit. 
We note that this upper limit is almost exactly an order of magnitude tighter than in \cite{Cyburt2005}.
For comparison, a scalar particle in equilibrium contributes $\Delta N_\nu = 4/7 = 0.57$,
so this is ruled out, unless it decouples well before neutrino decoupling \cite{Steigman:1979xp,Olive:1980wz}.

When the best fit value of $N_\nu$ is near (or as in Eq.~(\ref{eq:Nnu-tout}), below) 3, the resulting limit 2-$\sigma$ upper limit in Eq.~(\ref{eq:DeltaNnu2side}) maybe overly aggressive if we know that there are at least three weakly interacting neutrinos contributing to weak freeze-out. 
In this case, one can consider a weakened limit by normalizing the likelihood function using only $N_\nu \ge 3$ \cite{Olive:1995fi} to find the 1-sided limit
\begin{equation}
\frac{\int^{3 + \delta N_{\nu}}_3 {\cal L}(N_\nu) \ dN_\nu}{\int^{\infty}_3 {\cal L}(N_\nu) \ dN_\nu} = 0.9545
\label{int}
\end{equation}
Thus we demand here that $N_\nu > 3$ to accommodate the three known light neutrino species.
In that case, the last column of Table \ref{tab:etannu} gives 95.45\% CL values for 
$\delta N_\nu$, of which the strongest is
\begin{equation}
\label{eq:DeltaNnu3}
    \delta N_{\nu} = N_{\nu} - 3 \ < \ 0.226 
\end{equation}
based on the combination of CMB+BBN+$Y_p$+D/H.
This one-sided limit is weaker than the corresponding two-sided limit in Eq.~(\ref{eq:DeltaNnu2side}), as expected,
but again we see that a fully coupled scalar is ruled out.

The same constraint can be expressed in terms of limits
on the speed-up factor
\begin{equation}
    \xi  = \frac{H(N_{\nu})}{H_{\rm std}} 
    = \sqrt{1+ \frac{7 (N_\nu-3)}{43}}
\end{equation} 
Using Eq.~(\ref{eq:Nnu-tout}),
and propagating the error, we find
the $1\sigma$ range
\begin{equation}
\label{eq:speedup-lim}
    -0.020 < \xi-1 < 0.003    
\end{equation}
and the $2\sigma$ limits
\begin{eqnarray}
\xi-1 & < & 0.015  \ \ \ \ (\mbox{given 2-sided} \ \Delta N_\nu) \\
\xi_3-1 & < & 0.018  \ \ \ \ (\mbox{given 1-sided} \ \delta N_\nu )
\end{eqnarray}
which are notably larger than the $1\sigma$ upper limit
in Eq.~(\ref{eq:speedup-lim})
because the central $N_\nu$ value is below 3.
Here are throughout, the subscript 3 denotes 
this case of demanding $N_\nu \ge 3$
and thus using the one-sided limit.

Finally, we can express the excess energy density in terms of 
the critical density, i.e., we can find $\Omega_X$ for perturbation $X$.  
{\em During BBN}, the total energy is fully radiation dominated,
so that $\rho_{\rm tot} \approx \rho_{\rm rad}$ to high precision,
and thus at the start of the BBN epoch
(at 2 $\sigma$)
\begin{eqnarray}
    \left. \Omega_X \right|_{{\rm BBN}} & = & \frac{\rho_X}{\rho_{\rm rad}+\rho_X}
    = \frac{7 \Delta N_\nu/4}{43/4+7 \Delta N_\nu/4} < 0.028 
    \label{eq:OmegaX-lim-BBN}\\
    \left. \Omega_{X,3} \right|_{{\rm BBN}} & = & \frac{\rho_{X,3}}{\rho_{\rm rad}+\rho_{X,3}}
    = \frac{7 \delta N_\nu/4}{43/4+7 \delta N_\nu/4}
 <  0.035
    \label{eq:OmegaX-lim-BBN3}
\end{eqnarray}
where $\rho_{\rm rad}$ is the unperturbed radiation density.

By the present day, radiation is a small fraction of the total
energy density, for example with $\Omega_\gamma h^2 = 2.473 \times 10^{-5}$.
It is convenient to find the ratio of the density in $X$ to that of photons,
which at the start of BBN is
$(\rho_X/\rho_\gamma)_{\rm BBN} = g_X/2 \ (T_X/T)^4 = 7/8 \ \Delta N_\nu$.
If $X$ is decoupled by the start of BBN,
$T_X \propto 1/a$, and so today
$(\rho_X/\rho_\gamma)_0 = (\rho_X/\rho_\gamma)_{\rm BBN} (a_{\rm BBN} T_{\rm BBN}/a_0 T_0)^4$.
Using the conservation of comoving
entropy $S = 2\pi^2/15 \ g_{*S} T^3 a^3$,
we have $(aT)_0/(aT)_{\rm BBN} = (4/11)^{1/3}$ and so
\begin{eqnarray}
\left. \Omega_X \right|_{0}  &=&
\pfrac{\rho_X}{\rho_\gamma}
\Omega_{\gamma,0}  =
\pfrac{4}{11}^{4/3}
\pfrac{\rho_X}{\rho_\gamma}_{\rm BBN}    
\Omega_{\gamma,0} 
= 
\frac{7}{8}
\pfrac{4}{11}^{4/3}
\Omega_{\gamma,0} \
\Delta N_\nu  \\
&= & 5.6 \times 10^{-6} \ h^{-2} \ \Delta N_\nu
\end{eqnarray}
Thus today we have at $2\sigma$
\begin{eqnarray}
\left. \Omega_X \right|_{0} & < &
1.0 \times 10^{-6} h^{-2} \label{eq:OmegaX-lim}\\
\left. \Omega_{X,3} \right|_{0} & < &
1.3 \times 10^{-6} \, h^{-2} \ .
 \label{eq:OmegaX-lim-3}
\end{eqnarray}
This limit is much tighter than at BBN
simply because  the two epochs lie
squarely on opposite sides of matter-radiation equality.

\subsection{Applications: Constraints on New Physics Prior to BBN}

The limits on $N_\nu$ (Eq.~\ref{eq:Nnu-tout})
or equivalently the speedup factor (Eq.~\ref{eq:speedup-lim})
probe physics beyond the Standard Model where the only significant effect on BBN is
manifest through the expansion rate, and that $\eta$ is 
unchanged after BBN.
This covers a wide range of new physics possibilities
of ongoing interest.
Here we 
illustrate several examples, but we note that this is a necessarily incomplete sample of 
the large literature on this topic
(for reviews see \cite{Sarkar:1995dd,Cyburt2005,Jedamzik:2009uy,Pospelov:2010hj}).

\subsubsection{Right-Handed Neutrinos}

The simplest models generating non-zero neutrino masses require right-handed Standard Model singlet neutrinos. Often these states are quite massive as in the seesaw mechanism, and hence bear no effect on BBN. However, it is possible that light right-handed neutrinos (RHNs) or sterile neutrinos are present and contribute to the energy density at the time of BBN. The standard BBN treatment of neutrinos 
assumes the Standard Model: three generations of neutrinos with left-handed couplings to the $W$ and $Z$ bosons.  If other light ($m_\nu \ll 1\, \rm MeV$)
neutrinos exist
and are populated, these states would of course 
contribute to $N_\nu$.

In the case of Dirac
neutrinos, 
it is obvious from the limit in Eq.~(\ref{eq:DeltaNnu3}) that three RHNs with full weak-scale interactions (leading to $N_\nu = 6$) are badly excluded.  
However, RHN states 
are not efficiently populated by Standard Model interactions.  
Nevertheless it is possible, within the limits of either Eqs.~(\ref{eq:DeltaNnu2side}) or (\ref{eq:DeltaNnu3}), that three light right-handed states are present provided they decouple from the thermal bath sufficiently early. More precisely,
we can translate the limits on $N_\nu$ into limits on the RHN temperature at BBN, which can be related to their decoupling temperature and ultimately how strongly they couple to the SM. 

For example, for $N_{\nu_R}$ new states with mass $m_{\nu_R} \ll 1$ MeV and with temperature $T_{\nu_R} < T_{\nu_L}$, the energy density in relativistic states is
\beq
\rho_{R} = \rho_\gamma + \rho_e + \rho_{\nu_L} + \rho_{\nu_R} =  \frac{\pi^2}{30} \left(2 + \frac72 + \frac74 N_\nu +  \frac74 N_{\nu_R} \left(\frac{T_{\nu_R}}{T_{\nu_L}} \right)^4 \right) T^4 \, ,
\eeq
leading to a one-sided limit based on Eq.~(\ref{eq:DeltaNnu3}) of
\beq
\frac{T_{\nu_R}}{T_{\nu_L}} < 0.52,
\label{tnr}
\eeq
when we assume $N_{\nu_R} = 3$.
In order to obtain a right-handed neutrino temperature satisfying (\ref{tnr}),
the RHNs must decouple sufficiently early \cite{Steigman:1979xp,Olive:1980wz,Steigman:1986nh} so that the number of relativistic degrees of freedom at decoupling satisfies
\beq
g_{dR} > \frac{43}{4} \left( \frac{T_{\nu_L}}{T_{\nu_R}} \right)^3 = 75 \, .
\eeq
This in turn requires that the right-handed neutrinos decouple at a temperature $T_{d\nu_R} \gtrsim m_\tau$. Using a more precise calculation of the equation of state \cite{Srednicki:1988ce,Borsanyi:2016ksw}, we find that 
\beq
T_{d\nu_R} > 1.2~{\rm GeV}.
\eeq
Very generically, if we assume that right-handed neutrino decoupling is controlled by contact interactions mediated by a new $Z^\prime$ gauge boson then 
we are able to set a lower bound on the new gauge boson mass 
\beq
M_{Z^\prime} \gtrsim \left(\frac{T_{d{\nu_R}}}{T_{d{\nu_L}}} \right)^{3/4} \left(\frac{43/4}{g_{dR}} \right)^{1/8} \left(\frac{g'}{g}\right) M_Z  \simeq 8.7\left(\frac{g'}{g}\right) ~{\rm TeV} \, ,
\label{conslimZ}
\eeq
where $g'$ is the new gauge coupling, $g$ is the $SU(2)_L$ gauge coupling, and we have assumed $T_{d{\nu_L}} \sim 2$ MeV \cite{Dicus:1982bz}.

The limit (\ref{conslimZ}) is conservative in the sense that we have used the one-sided limit on $N_\nu$ requiring $N_\nu \ge 3$. A significantly stronger limit on $M_{Z^\prime}$ is possible if we employ the more restrictive two-sided limit from Eq.~(\ref{eq:DeltaNnu2side}).
In that case, we find $T_{\nu R}/T_{\nu L} < 0.49$, which gives $g_{\nu R} > 89$.
This is only possible if $T_{d{\nu_R}} > M_Z$, or more accurately from \cite{Borsanyi:2016ksw}, $T_{d{\nu_R}} > 30$~GeV. 
This gives a limit $M_{Z^\prime} \gtrsim 95$~TeV. Some model-specific limits using the arguments above can be found in Refs.~\cite{Steigman:1986nh,Lopez:1989dh,Faraggi:1990ita,Gonzalez-Garcia:1989ygi,Barger:2003zh,Solaguren-Beascoa:2012wpm}.  These limits can be compared to experimental limits on extra $U(1)$ gauge bosons, which are necessarily model-dependent as they depend on how these gauge bosons are coupled to Standard Model particles.  For example \cite{2020PTEP.2020h3C01P}, limits on $Z_{LR}$ in a left-right symmetric model are
$M_{Z_{LR}} \gtrsim 1.2$~TeV and mass limits on a $Z^\prime$ gauge boson with SM couplings are $M_{Z^\prime} > 5.1$ TeV. 

This generic argument assumes that thermalization occurs through contact interactions below the scale of the gauge boson mass. The derivation of the limit is more complicated at smaller values of the $Z'$ gauge coupling where thermal decoupling instead occurs for $T\sim m_{Z'}$ through $Z'$ decays and inverse decays, while at even smaller couplings out-of-equilibrium freeze-in production can still be constraining. Such effects were worked out for the case of a light, feebly-coupled $U(1)_{B-L}$ gauge boson in~\cite{Abazajian:2019oqj, Adshead:2022ovo}; for $1\,\mathrm{MeV}\lesssim m_{Z'} \lesssim $ 100 GeV and couplings $g'\ll 1$, the resulting contribution of RH neutrinos to $\Delta N_\nu$ constitutes the leading probe of the model, together with limits from Supernova 1987A.

Alternatively, the degree to which sterile neutrinos are
present in the thermal bath \cite{Enqvist:1990ad,Enqvist:1990ek,Enqvist:1996bj,Dolgov:2003sg,Asaka:2006rw,Asaka:2006nq} may be constrained by the possible mixing with Standard Model left-handed neutrinos. For example, if $\theta_s$ is defined as the mixing angle between active and sterile neutrinos,
limits on $\theta_s$ can be derived. If $\nu_1 \approx \cos \theta_s \nu_s + \sin \theta_s \nu_L$ and $\nu_2 \approx \cos \theta_s \nu_L - \sin \theta_s \nu_s$ are the two mass eigenstates, 
the interaction rates for $\nu_1$ are similar to those of the active neutrinos, $\nu_2$, though suppressed by $\theta_s^2$ for $\theta_s \ll 1$. As in the examples discussed above, $\nu_1$ will decouple before $\nu_2$ with $T_{d1} \sim T_{d2} \theta_s^{-2/3}$ and the limits on $T_{\nu_R}$ can be converted into a limit on $\theta_s$ \cite{Dudas:2020sbq}.
Assuming a single sterile state, $T_1/T_2 < 0.69$
requiring $T_{d1} \gtrsim \Lambda_{\rm QCD}$
\beq
\theta_s < \left( \frac{T_{d2}}{T_{d1}} \right)^{3/2} = \left( \frac{2 {\rm MeV}}{\Lambda_{\rm QCD}} \right)^{3/2} \lesssim 1.5 \times 10^{-3} \, ,
\eeq
for $\Lambda_{\rm QCD} = 150$ MeV. 
Other cosmological and astrophysical limits have also been recently discussed \cite{Alonso-Alvarez:2022uxp}.

Finally, right-handed neutrino states
can allow for the presence of neutrino magnetic moments, whose cosmological effects depends on whether the neutrinos are Dirac or Majorana.  In either case, a complete analysis includes changes in neutrino energy spectra and in their cosmic thermodynamics, which are beyond the scope of our analysis;
see refs.~\cite{Morgan:1981zy,Grasso:1996yx,Vassh2015,Balantekin2018} and references therein.

\subsubsection{Dark Radiation}

Thermal dark radiation is a frequent ingredient in theories of beyond-the-SM physics.  
Pseudo-Nambu-Goldstone bosons (pNGBs), for instance, appear in many extensions of the SM, arising from the spontaneous breaking of (for instance) Peccei-Quinn symmetries (axions) \cite{Weinberg:1977ma,Wilczek:1977pj}; lepton number symmetries (majorons) \cite{Chikashige:1980ui,Gelmini:1980re}; or family symmetries (familons) \cite{Wilczek:1982rv}.   PNGBs are naturally light and, depending on the reheating temperature and the symmetry-breaking scale, may be thermally populated in the early universe, thereby contributing to dark radiation.  However, stellar and supernova cooling bounds on the interactions of these pNGBs with matter typically require that the symmetry breaking scale is sufficiently large to preclude the pNGB from being in equilibrium with the SM plasma at temperatures below the QCD phase transition \cite{Brust:2013ova}. Thus in minimal models pNGBs generically do not lead to shifts in $\Delta N_{\rm eff}$ that are large enough to be detected with current sensitivity; however, they represent well-motivated targets for the future.  In less minimal models, such as those that contain also a dark matter candidate, e.g.~\cite{Weinberg:2013kea}, stellar cooling bounds can be relaxed and substantially larger contributions to $\Neff$ are possible.

More generally,  
multi-state hidden sectors offer many avenues to address the origin of dark matter (DM), whether DM is a thermal or non-thermal relic. DM may be produced by a thermal freeze-out process from a dark radiation bath, stabilized through a dark number asymmetry similar to the baryon asymmetry, or through a more complicated interplay of equilibrium and non-equilibrium mechanisms; see e.g.~\cite{Asadi:2022njl} for an overview of recent work.
In such hidden sector scenarios, dark matter is produced out of a dark thermal bath, which carries a 
sizeable amount of entropy.  The simplest way to accommodate this entropy is to sequester it in dark radiation, which requires the existence of a light degree of freedom in the hidden sector spectrum.  Otherwise the lightest hidden sector state(s) must decay into the SM in the early universe, which imposes severe restrictions on the abundance and lifetime of the decaying particle(s).  A relativistic relic contributing to dark radiation thus represents a generic (though not universal) component of dark sector model building.   
 
Thermal hidden sectors in the early universe are also motivated by approaches to the hierarchy problem.  
The Twin Higgs mechanism introduces a mirror or `twin' copy of the SM, related to the SM by a discrete symmetry that ensures a cancellation between the contributions of SM and twin particles to the Higgs mass parameter \cite{Chacko:2005pe}.   Exact cancellations require the introduction of twin photons and twin neutrinos, which can all contribute to dark radiation. Given the Higgs portal interactions inherent between twin and SM particles, twin particles are  in thermal equilibrium with the SM in the early universe until temperatures of $\mathcal{O}(\mathrm{GeV})$, and thus mirror Twin Higgs models require a period of late asymmetric reheating in order to dilute the dark radiation to levels allowed by current constraints \cite{Chacko:2016hvu, Craig:2016lyx}.
Another approach to the hierarchy problem is furnished by NNaturalness \cite{Arkani-Hamed:2016rle}, which postulates a large number $N$ of copies of the SM that differ by the value of the Higgs mass-squared parameter, with Higgs-reheaton couplings arranged such that the sector with the smallest non-zero Higgs vacuum expectation value is preferentially populated in the early universe.  The massless species from the additional sectors, populated at much lower temperatures than the SM, then contribute to dark radiation.  

In a self-interacting dark sector, the dark radiation may have cosmologically relevant interactions with other dark species.  In some theories these interactions can keep the dark radiation in internal kinetic equilibrium, so that it acts as a perfect fluid, rather than a free-streaming relic, during recombination.  Both fluid and free-streaming dark radiation affect BBN in the same way, through the contribution of their (homogeneous) energy density to the Hubble rate, and therefore both are subject to BBN constraints on $N_\nu$. However, the imprint of fluid dark radiation on the power spectrum of the CMB anisotropies has a different phase than in the free-streaming case appropriate for SM neutrinos, and the contribution of fluid dark radiation must be  quantified by the observable $N_{\mathrm{fluid}}$  rather than $N_{\mathrm{eff}}$ \cite{Chacko:2003dt,Bell:2005dr}.  CMB constraints on $N_{\mathrm{fluid}}$ are typically weaker by a factor of 2-3 than the corresponding CMB constraints on $N_\mathrm{eff}$ \cite{Brust:2017nmv,Blinov:2020hmc}. Theories with fluid dark radiation are still subject to the BBN-only constraints on $N_\mathrm{eff}$ derived in Section~\ref{sect:indepdendent}.

The energy density carried in thermal dark radiation depends on its temperature $T_{DR}$, which will in general differ from that of the SM, as well as the number of its degrees of freedom $g_{*HS}$.  Dark radiation contributes to the energy density in relativistic species as 
\beq
\label{eq:dr}
\rho_{\rm DR} = g_{*\rm HS}(T_{\rm DR})\frac{\pi^2}{30} T_{\rm DR}^4. 
\eeq 
In describing multi-state hidden sectors, we need to account for a possible time-varying $g_{*\rm HS}$ as hidden sector species may become nonrelativistic and deposit their entropy into the remaining dark radiation.  Denoting the value of $g_{*\rm HS}$ immediately prior to BBN as $g_{*\rm HS}^{\rm IR}$, from Eq.~(\ref{nnutxtnu}) the contribution to
$\Delta N_\nu$ is
\beq
\Delta N_\nu = \frac{4}{7} g_{*\rm HS}^{\rm IR} \left(\frac{T_{\rm DR}}{T_{\nu}} \right)^4 = 2.2  g_{*\rm HS}^{\rm IR} \left(\frac{T_{\rm DR} }{T_{\gamma}}  \right)^4, 
\eeq 
where in the last equality we have used $T_\nu = (4/11)^{1/3}T_\gamma$.  Assuming entropy conservation, we can rewrite this expression in terms of the HS and SM temperatures and effective entropic degrees of freedom $g_{*S} (T)$ at some earlier time as
\beq
\Delta N_\nu = 2.2  g_{*HS}^{IR}  \left(\frac{T_{DR}^{UV} }{T_{\gamma}^{UV}}  \right)^4 \left(\frac{43/4}{g_{*S,SM}(T_{\gamma}^{UV})} \right)^{4/3}  \left( \frac{g_{*S,HS}(T_{DR}^{UV})}{g_{*S,HS}(T_{DR}^{IR})} \right)^{4/3}.
\eeq
Thus if the HS and the SM were in thermal equilibrium at some common UV temperature $T_\gamma^{\rm UV} = T_{\rm DR}^{\rm UV} = T_*$, the resulting shift in $N_\nu$ depends only on the evolution of the numbers of degrees of freedom in both sectors following decoupling:
\beq
\label{eq:decouplingdr}
\Delta N_\nu = 2.2  g_{*HS}^{IR} \left(\frac{43/4}{g_{*S,\rm SM}(T_{*} )}\right)^{4/3}  \left( \frac{g_{*S,\rm HS}(T_{*})}{g_{*S,\rm HS}(T_{\rm DR}^{\rm IR})} \right)^{4/3} . \eeq
The one-sided upper bound from Eq.~(\ref{eq:DeltaNnu3}) requiring $\delta N_\nu < 0.226$ thus allows two relativistic degrees of freedom to have been in equilibrium with the SM at early times (i.e., taking $g_{*S,\rm SM} = g_{*,\rm SM} = 106.75$).  This bound is stringent enough to preclude nontrivial evolution in $g_{*\rm HS}$ once the HS has decoupled from the SM: e.g., putting $g_{*S,\rm HS}(T_{*}) =2$ and  $g_{*\rm HS}^{\rm IR} = g_{*S,\rm HS}^{\rm IR} = 1$ in Eq.~(\ref{eq:decouplingdr}) results in $\Delta N_\nu = 0.26$.  Of course, if there are additional BSM species in the thermal plasma at early times that deposit their entropy preferentially into the SM, the temperature of the dark radiation is then further suppressed relative to that of the SM, and more degrees of freedom can be accommodated.  Dark sectors may also be populated out of equilibrium with the SM in the early universe, e.g.~via asymmetric reheating \cite{Adshead:2016xxj}, in which case the temperature ratio between the photons and the dark radiation is a free parameter and $g_{*\rm HS}$ may be substantially larger.

\subsubsection{Stochastic Gravitational Wave Background}

A gravitational wave background is a generic prediction of inflation models.  After inflation, processes in the early universe such as phase transitions or the evolution of cosmic string networks can also source stochastic gravitational wave backgrounds; see, e.g., ref.~\cite{Maggiore:1999vm}.
Regardless of its origin,
the energy density in gravitational waves laid down before BBN
acts as radiation and thus its impact on BBN is completely
captured by our $N_\nu$ analysis.  Gravitational waves with adiabatic initial conditions leave the same imprint on the CMB as free-streaming dark radiation, and for this case 
the limit on the present-day energy density in gravitational
waves is that in Eq.~(\ref{eq:OmegaX-lim-3}), 
namely
\begin{equation}
\label{eq:GWBG}
    \Omega_{\rm GW,0} \, h^2 < 1.3 \times 10^{-6}
    \ \ \ (f \gtrsim 2 \times 10^{-11} \ {\rm Hz})
\end{equation}
where we impose $N_\nu > 3$.
This is similar to recent ``indirect'' limits
based on BBN and the CMB and 
quoted by the LIGO and VIRGO collaborations \cite{LIGOScientific:2016jlg,KAGRA:2021kbb}.
Other recent limits using the CMB,
often in concert with BBN,
are in refs.~\cite{Pagano:2015hma,Lasky:2015lej,Li:2016mmc,Li:2021htg}.
Gravitational waves with non-adiabatic initial conditions give rise to a different CMB signature and can be more tightly constrained; results appear in
ref.~\cite{Clarke:2020bil}.

The BBN constraint on stochastic gravitational
waves applies to frequencies above a cutoff $f \gtrsim f_{\rm BBN}$. This restriction on the frequency range arises 
because the BBN limit applies to modes
that are within the horizon at the start of BBN,
so that the comoving wavenumber is larger than the inverse of the comoving Hubble length $d_{\rm H,com} = 1/aH$ at the time, i.e.,
$k_{\rm BBN} > a_{\rm BBN} H_{\rm BBN}$ \cite{Boyle:2007zx}.
Setting $T_{\rm BBN} = 1 \, \rm MeV$ gives $z_{\rm BBN} = 5.9 \times 10^9$,
and then $f_{\rm BBN} = k_{\rm BBN}/2\pi = 1.8 \times 10^{-11} \ \rm Hz$.
A similar argument gives a CMB frequency cutoff $f_{\rm CMB} \gtrsim 10^{-16} \ \rm Hz$.
This means that while the limit in Eq.~(\ref{eq:GWBG}) applies for the frequency range indicated,
if there is a gravitational wave component with $f_{\rm CMB} < f < f_{\rm BBN}$,
this will contribute to $N_\nu^{\rm CMB}$ but not to $N_\nu^{\rm BBN}$.
This would lead to an effective time variation in $N_\nu$, which is the subject of \S \ref{sect:difference}.

\subsubsection{Vacuum Energy: Tracker Solutions}

A scalar field $\phi$ present during BBN
contributes vacuum energy that affects
the cosmic expansion rate
\cite{Sarkar:1995dd}. 
In particular, 
quintessence models for dark energy 
often have tracker solutions,
in which the scalar field providing
the dark energy behaves as
the dominant cosmic mass-energy  component
until late times when the field drives cosmic acceleration today
\cite{Sarkar:1995dd,Birkel:1996py,Ferreira:1997hj,Bean:2001wt,Kneller2003}.  Thus, $\rho_\phi$ effectively acts as a subdominant component of radiation during BBN,
and is amenable to our analysis.

The limit applies during BBN, where
the energy density must obey
Eq.~(\ref{eq:OmegaX-lim-BBN3}), i.e.,
\begin{equation}
\left. \Omega_{\phi}\right|_{\rm BBN} < 0.035   
\end{equation}
Note that for this result is valid 
when the tracker field
behaves like radiation throughout BBN
and CMB, with the same fraction of 
the radiation density over all of this time.

For a potential $V(\phi) = \mpl^4 e^{-\lambda \phi/\mpl}$, the coupling is constrained to
be $\lambda = 2\mpl/\sqrt{\Omega_\phi} > 11$.

\subsubsection{Changing Fundamental Couplings}

Though there  is no intrinsic reason that fundamental couplings are in fact constant,
there are many constraints which limit their time variation. For a review see: \cite{Uzan:2010pm}. Some of the strongest limits are derived from BBN since the temporal baseline is essentially the age of the Universe. In most cases, the limits derived from BBN can be traced to their effect on the freeze-out of the weak interactions which determine the neutron-to-proton ratio. That is, any new physics which alters either Eqs.~(\ref{Gweak}) or (\ref{H}), will affect $T_f$ and hence $n/p$.
Very simply, we have 
\beq
\frac{\Delta (n/p)}{(n/p)} = \frac{\Delta m}{T_f} \frac{\Delta T_f}{T_f} \, ,
\label{dnp}
\eeq
which induces a change in the light elements, most notably $Y_p \approx 2(n/p)/(1+(n/p))$. 
This is the basis for the limits on $N_\nu$ and more generally $\xi$ as these directly affect $H$ in Eq.~(\ref{H}). 

A variation in the fine-structure constant, $\alpha$, however affects the neutron-proton mass difference (which then affects $n/p$). 
The neutron-proton mass difference receives both weak and electromagnetic contributions
\beq 
\Delta m =  a \, \alpha \, \Lambda + b (h_d-h_u) \,v \,,
\label{dm}
\eeq 
where the electromagnetic contribution is $a \, \alpha \, \Lambda = -0.76 \ \rm MeV$ and is proportional to the QCD scale $\Lambda$, while the 
weak contribution is $b ({h_d}-{h_u})\, v = 2.05 \ \rm MeV$, where $h_{d,u}$ are the Yukawa couplings to the $u$ and $d$ quarks and $v$ is the Higgs expectation value \cite{Gasser:1982ap}. Therefore a change in $\alpha$ directly affects $\Delta m$
and Eq.~(\ref{dnp}) is modified
\beq
\frac{\Delta (n/p)}{(n/p)} = \frac{\Delta m}{T_f} \left(\frac{\Delta T_f}{T_f} - \frac{\Delta^2 m}{\Delta m} \right) \, ,
\label{dnp2}
\eeq
where $\Delta^2 m$ is the change in $\Delta m$.
This leads to a variation in $Y_p$
\beq
\frac{\Delta Y_p}{Y_p} = \frac{\Delta m/T_f}{1+(n/p)} \left(\frac{\Delta T_f}{T_f} - \frac{\Delta^2 m}{\Delta m} \right) \approx 1.3 \frac{\Delta^2 m}{\Delta m} \, ,
\eeq
where the last inequality assumes $T_f \approx 0.84 $ MeV. Thus the bounds on $Y_p$ constrain $\Delta^2 m$ and hence $\alpha$ \cite{Kolb:1985sj,Campbell:1994bf,Bergstrom:1999wm,Ichikawa:2002bt,Nollett:2002da}.

From Eq.~(\ref{dm}), $\Delta^2 m = -0.76 \Delta \alpha/\alpha \ \rm MeV$ and using the observational uncertainty in $Y_p$ from Eq.~(\ref{eq:Ypobs}), 
we have
\beq
\frac{\Delta \alpha}{\alpha} \lesssim 0.018 \, .
\label{alimit}
\eeq
Though it probes a different cosmological epoch,
Planck constraints on variations in $\alpha$ from the CMB are now tighter than that from BBN.
Intermediate Planck results \cite{Planck:2014ylh} give
$\Delta \alpha/\alpha =  (3.6 \pm 3.7) \times 10^{-3}$ or $\Delta \alpha/\alpha <  0.007$ at 1 $\sigma$, somewhat stronger than the limit in Eq.~(\ref{alimit}). 
However, in many theories which allow for a variation in $\alpha$, that variation is accompanied by variations in other fundamental parameters \cite{Campbell:1994bf,Langacker:2001td,Dent:2001ga,Calmet:2001nu,Dent:2003dk,Calmet:2006sc} affecting not only the neutron-proton mass difference, but also the deuterium binding energy and other quantities relevant to BBN \cite{Campbell:1994bf,Ichikawa:2002bt,Muller:2004gu,Flambaum:2002de,Dmitriev:2003qq,Landau:2004rj,Coc:2006sx,Dent:2007zu,Coc:2012xk,Martins:2020syb,Deal:2021kjs}.  These considerations are more model dependent, but generally provide limits which are roughly two orders of magnitude stronger than (\ref{alimit}). 

While variations in $\alpha$ and other couplings affect the weak interaction rates in Eq.~(\ref{Gweak}), a variation in the gravitational constant directly affects Eq.~(\ref{H}) and the limit on $G_{\rm N}$ can easily be cast in the form of a limit on $N_\nu$. Of course any consideration of a variation in $G_{\rm N}$,
implicitly assumes the constancy of some masses, e.g., the proton mass, so that it is the variation of the gravitational coupling $G_{\rm N} m_p^2$ which is constrained. 

From Eq.~(\ref{H}), we see that $H^2 \propto G_{\rm N}$,
and we thus have
$\Delta G_{\rm N}/G_{\rm N} = \xi^2-1 = (7/43) (N_\nu - 3)$. Using $N_\nu = 2.898 \pm 0.141$
we find the 1 $\sigma$ range 
\begin{equation}
-0.040 < \frac{\Delta G_{\rm N}}{G_{\rm N}}
< 0.006 \  \ .
\end{equation}
The asymmetry in the bounds
is traced to the
central value of $N_\nu$ lying
sightly  below 3.
More conservatively we have at 2 $\sigma$
\begin{eqnarray}
-.062 < \pfrac{\Delta G_{\rm N}}{G_{\rm N}}
& < & 0.029 \\
0 < \pfrac{\Delta G_{\rm N}}{G_{\rm N}}_3
& < & 0.037 \, 
\end{eqnarray}
which is similar to recent work \cite{Alvey:2019ctk}.
If we parameterize the variation as $G_{\rm N} \sim t^{-x}$
\cite{Yang:1978ge,Accetta:1990au,Copi:2003xd,Umezu:2005ee,Bambi:2005fi,Alvey:2019ctk},
the 2 sigma limits become $-0.002 < x < 0.0007$
and $x < 0.0009$, respectively. These corresponds to a limit $-5.1 \times 10^{-14}~{\rm yr}^{-1} < {\dot G}_{\rm N}/G_{\rm N} < 1.1 \times 10^{-13}~ {\rm yr}^{-1}$
and $-6.5 \times 10^{-14}~{\rm yr}^{-1} < {\dot G}_{\rm N}/G_{\rm N} < 0$ (for the 2-sided and 1-sided limits), which is slightly better than the strongest limit from lunar-laser-ranging \cite{Hofmann:2018myc} giving ${\dot G}_{\rm N}/G_{\rm N} < 2.2 \times 10^{-13}$ yr$^{-1}$.

As a variation in $G_{\rm N}$ effectively implies a non-minimal theory of gravity, we can translate the bounds on the variation into bounds on parameters in specific non-minimal models.
For example, in Brans-Dicke gravity, the variation in $G_{\rm N}$ can be translated in a limit on the coupling $\omega$ ($\omega \to \infty$ corresponds to the limit of Einstein gravity) which is bound by BBN \cite{Casas:1990fz,Casas:1991ui,1992ApJ...391..433S,Clifton:2005xr,Umezu:2005ee}. Our 2-sided and 1-sided limits on $\omega$ are $\omega > 270$ and $> 230$ respectively. These are weaker than bounds imposed by the CMB which are of order 2000
\cite{Chen:1999qh,Nagata:2003qn,Ooba:2016slp,Ooba:2017gyn}.
Both cosmological limits are weaker than the 
limits obtained by Doppler tracking of the Cassini spacecraft \cite{Bertotti:2003rm} which gives $\omega > 15000$  (2 $\sigma$) \cite{Will:2014kxa}. 

\subsubsection{Primordial Magnetism}

Primordial magnetic fields could be created
in the very early universe, e.g., during inflation. The fields would thus be present
during BBN.
Jedamzik et al \cite{Jedamzik1998}
showed that magnetic field modes
are damped at scales much below the 
horizon at neutrino decoupling.  
As Cheng et al \cite{Cheng1996} note,
this means that magnetic fields during BBN
are well-approximated as a uniform 
perturbation to the energy density.
The energy density
$\rho_{B} = B^2/8\pi \propto a^{-4}$ scales as
radiation because the field obeys
$B \propto a^{-2}$ due to flux conservation.

Magnetic fields during BBN not only 
add to the energy density through $\rho_B$.
They also perturb the $e^\pm$ density of states,
and thus (1) boost the
the energy density and pressure due
to pairs, and (2) change the $n \rightarrow p$
interconversion rates.  
For field strengths of interest,
the effect of $\rho_B$  dominates
the perturbation to BBN, and thus
is amenable to our treatment;
see refs.~\cite{Cheng1996,Kernan1996,Kawasaki2012} for detailed analysis.

Taking the approximation of the field's
energy density as the dominant perturbation,
we have
$B^2/8\pi < \Delta N_\nu \ \rho_{1\nu}$.
At the beginning of BBN, this corresponds to
$B(1 \ {\rm MeV}) < 2.6 \times 10^{13} \ \rm Gauss$;
today
this would be $B_0 < 2.8 \ \rm \mu G$.

\section{Searching for new physics between the BBN and CMB epochs}
\label{sect:difference}

Up until now,  we have assumed that the values of $\eta$ and $N_\nu$ remained unchanged between the time of BBN and CMB decoupling. However, it is possible that new physics is responsible for a change in these quantities and therefore,  
we now drop the assumption that one or both of $N_\nu$
and $\eta$ are the same for the BBN and CMB epochs.
This probes possible evolution in these quantities.
The data allow us to probe 
scenarios where only one of $\eta$
or $N_\nu$ change, or where
both change.  As we will
see, these correspond to different
physical scenarios.
For example, a stable particle becoming
non-relativistic after BBN would change
$N_\nu$ only, while decays or annihilations of some beyond the Standard Model particles into Standard Model  particles
can change both $\eta$ and $N_\nu$.

The baryon-to-photon ratio is intimately related to the comoving entropy
of relativistic species
\beq
S_{\rm com} = a^3 \, s =
\frac{2 \pi^2}{45} \ g_{*,S} \ (a T)^3 \ \  .
\eeq
where $s$ is the entropy density and $a$ is the cosmological scale factor.
Here $g_{*,S} = \sum_{\rm boson} g_b (T_b/T)^3
+ 7/8 \ \sum_{\rm fermion} g_f (T_f/T)^3$ is
the usual effective number of entropic degrees of freedom, summed
over relativistic species.
For constant $g_{*,S}$, we see that $S_{\rm com} \propto (aT)^3$, and thus
the photon temperature $T \propto 1/a$ when $S_{\rm com}$ is conserved. Note also that the baryon density (assuming baryon number conservation) scales as $n_B \propto 1/a^3$, and the photon number density $n_\gamma \propto T^3$, 
the baryon-to-photon ratio can be normalized to its
present value $\eta_0$
\beq
\eta = \frac{n_{\rm B}}{n_\gamma}  = \eta_0 \pfrac{a_0 T_0}{aT}^3 \, .
\eeq
Thus a change in entropy and a change in the product $aT$
can be related to $\eta$
\beq
\label{eq:eta-entropy}
\eta = \frac{2 \pi^2 g_{*,S} a_0^3 T_0^3 \eta_0}{45 S_{\rm com}}   \ \ .
\eeq
Thus we have $\eta \propto (g_{*,S}/S_{\rm com})$,
and is it clear that entropy
production---changing the comoving entropy $S_{\rm com}$---leads to changes in $\eta$.

We note that variations 
in $\eta$ can in some cases be linked to changes in $N_\nu$.
This is because the cosmic entropy is dominated by that in relativistic species via the second law
of thermodynamics:  $T \, dS_{\rm com} = d(\rho_{\rm rel} a^3) + p_{\rm rel}d(a^3) = a^{-1} d(\rho_{\rm rel} a^4)$,
where $\rho_{\rm rel}$ and $p_{\rm rel}$ correspond to the energy density and pressure of relativistic species.
Thus entropy change $dS_{\rm com} \ne 0$ 
in relativistic species
also implies
a change in $\rho a^4$, which can be parameterized 
by $N_\nu$.

We therefore treat $\eta^{\rm BBN}$ and $\eta^{\rm CMB}$ 
as distinct, as we do for $N_\nu^{\rm BBN}$ and $N_\nu^{\rm CMB}$.
To probe their joint distribution, we
convolve over the light element abundances
\begin{eqnarray}
\label{eq:4D}
{\cal L}(N_\nu^{\rm BBN},\eta^{\rm BBN},N_\nu^{\rm CMB}, \eta^{\rm CMB}) &  & \\
\nonumber
& \propto  & 
\int {\cal L}_{\rm NCMB}(\eta^{\rm CMB},N_\nu^{\rm CMB},Y_p) 
\ {\cal L}_{\rm NBBN}(\vec{X};\eta^{\rm BBN}, N_\nu^{\rm BBN}) 
\prod_i {\cal L}_{\rm obs}(X_i) \ dX_i \ \ .
\end{eqnarray}
This serves to link the CMB and BBN distributions
via their dependence on $Y_p$.  

We continue to assume 
the light elements did not change after BBN,
so that the their observations are valid to apply for
both BBN and the CMB.
In particular, by using low-redshift  astronomical observations 
$Y_{p,\rm obs}$ from extragalactic HII regions
(Eq.~\ref{eq:Ypobs}),
we are implicitly assuming that there is not any cosmologically important (i.e., pre-stellar) nucleosynthesis
between BBN and recombination:
\begin{equation}
\label{eq:He-equality}
Y_p^{\rm BBN} = Y_p^{\rm CMB} = Y_{p,\rm obs}  \ \ .
\end{equation}

There are several
ways to proceed to understand the constraints on this 4-dimensional space.  
We first examine possible changes in $N_\nu$ alone.

\subsection{Limits on $\mathrm{N_\nu}$ Evolution}

Our evaluations of $N_\nu$ in Fig.~\ref{fig:N_nu_dist_Aver} include
both CMB-only and BBN-only analyses,
which give independent measures.  
We see that the two 
measures are remarkably consistent, $|N_\nu^{\rm CMB} - N_\nu^{\rm BBN}|$ is roughly one third of the half-width of the distributions.
We already see that
the CMB and BBN show no {\em need} for a change in $N$
between the epochs, a theme we will see repeated.  Thus it is reasonable to combine them as we
have in the previous section; here we 
examine the extent to which they are allowed to
differ. Processes which may change the value of $N_\nu$ between BBN and CMB decoupling include decays of non-relativistic particles to relativistic ones which would increase $N_\nu$, or some relatively light particle (say of order 1 keV), becoming non-relativistic, thereby decreasing $N_\nu$.

We marginalize over the two $\eta^i$ to compare the neutrino numbers and produce the likelihood
\beq
\label{eq:indep-eta}
{\cal L}(N_\nu^{\rm BBN},N_\nu^{\rm CMB})
\ \propto \ \int {\cal L}(N_\nu^{\rm BBN},\eta^{\rm BBN},N_\nu^{\rm CMB}, \eta^{\rm CMB}) 
\  d\eta^{\rm BBN} \ d \eta^{\rm CMB} \, .
\eeq
In the marginalization, we adopt the least restrictive assumption regarding $\eta^i$, 
 i.e.,  we do not demand that the two $\eta^i$ are related, and allow them to differ.
This is appropriate for scenarios where a change in $\eta$
is possible.

\begin{figure}
    \centering
    \includegraphics[width=\textwidth]{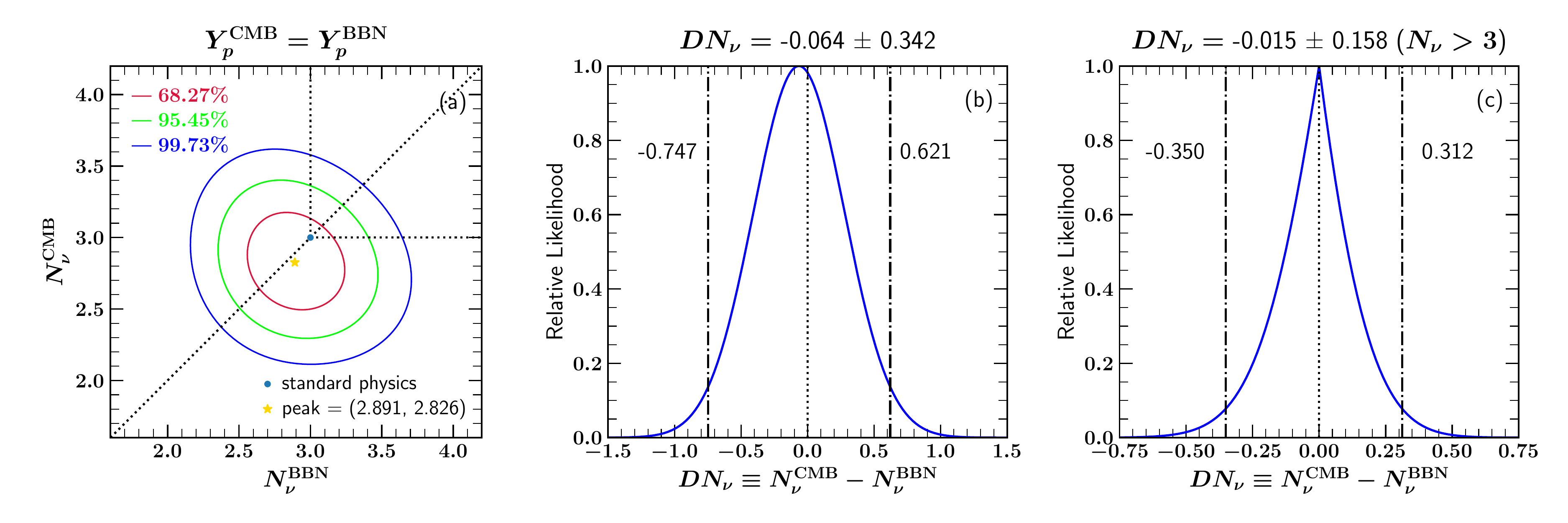}
    \caption{CMB vs BBN independent determinations of $N_\nu$.  We marginalize over $\eta$ in both epochs separately as shown in Eq.~(\ref{eq:indep-eta}), after convolving with light elements as in Eqs.~(\ref{eq:4D}) and~(\ref{eq:He-equality}).  We adopt the Aver et al~$Y_p$ \cite{aver2021} as in Eq.~(\ref{eq:Ypobs}).
    {\em Panel (a)} compares the two measures, showing that the best-fit point is quite close to the diagonal $N_\nu^{\rm CMB} = N_\nu^{\rm BBN} $ line, and within 1 sigma of the Standard Cosmology result at $(3,3)$. We note that the BBN and CMB constraints are quite comparable to each other.  {\em Panel (b)} shows the distribution of the {\em difference} $D N_\nu \equiv N_\nu^{\rm CMB}-N_\nu^{\rm BBN}$, which peaks nearly at the Standard Cosmology value $\Delta N_\nu = 0$. The vertical dot-dashed lines show the 95\% two-sided limit.
    {\em Panel (c)} is similar to panel (b) but requiring $N_\nu \ge 3$.
    Vertical lines give the 95\% one-sided limits. Overall we see that our results are in excellent agreement with the Standard Cosmology expectations, and thus we place limits on nonstandard alternatives where $\Delta N_\nu \ne 0$.}
    \label{fig:Nnu_compare_same_Yp}
\end{figure}

Results for Eq.~(\ref{eq:indep-eta})
appear in Fig.~\ref{fig:Nnu_compare_same_Yp}.
Panel (a) shows the joint constraints  
in the $(N_\nu^{\rm BBN},N_\nu^{\rm CMB})$ plane.
We we see that the $1\sigma$ contour
includes the Standard Model case of $(3,3)$,
while the best fit likelihood peak is slightly below 3
in both variables.  The dotted diagonal line 
shows $N_\nu^{\rm CMB} = N_\nu^{\rm BBN}$,
which almost perfectly bisects the contours, and the best-fit point (shown by the yellow dot) is very close to this line.
Indeed, the nearly circular contours also show that the BBN and CMB 
constraints have very similar spreads
(and that the $Y_p$ convolution does not introduce
much correlation).
These results reaffirm
that the Standard Model works
very well:  neither BBN nor the CMB provide any
significant evidence for new physics.
Moreover, the remarkable agreement
between the independent values of $N_\nu$
measured by the CMB and BBN also
shows that there is no preference for 
a change in $N_\nu$ between the two epochs.
Thus we proceed to use these results to limit
changes in $N_\nu$.

Panel (b) of Fig.~\ref{fig:Nnu_compare_same_Yp}
explores the allowed range in the
\begin{equation}
D N_\nu \equiv N_\nu^{\rm CMB}-N_\nu^{\rm BBN}    
\end{equation}
difference. Here for each value of $D N_\nu$,
we marginalize the distribution in panel (a) to obtain the likelihood, ${\cal L}(D N_\nu)$ shown in panel (b).
We see that the peak is very near zero (the Standard Model value is shown by the vertical dotted line),
indicating that there is no preference for 
$N_\nu$ evolving between these two epochs.
The best fit being $D N_\nu = -0.064 \pm 0.342$.  Note that the uncertainty here is related to the uncertainties obtained in the previous section for BBN-alone and CMB-alone, and in this case as one might expect, the combined uncertainty in $D N_\nu$ is larger than either of the two individually. 
The vertical dot-dashed lines give the 95\% 2-sided limits on $D N_\nu$.  We  see that while the $2\sigma$ limits on the difference in $N_\nu$
do not allow 1 full neutrino species, but
could allow a net gain or loss of a fully-coupled scalar.

Panel (c) of Fig.~\ref{fig:Nnu_compare_same_Yp}
is similar to panel (b), 
but now requires that both the BBN and CMB values
have $N_\nu >3$.
This amounts to marginalizing over fixed values of $D N_\nu$ within
the top right corner outlined by dotted lines in panel (a).
The distribution in panel (c) peaks sharply at $D N_\nu \simeq 0$, indicating consistency with the 
Standard Model.\footnote{
The cuspy nature of the curve partially
reflects the sharp boundaries imposed
by demanding $N_\nu > 3$,
but even so the peak need not occur at $D N_\nu = 0$.  The peak would be offset from zero
if the peak likelihood in panel (a) were
roughly more than $1\sigma$ away from the diagonal.}
Vertical dot-dashed lines show 95\% 
one-sided limits on the difference, which we
see are quite strong:  
positive values have $D N_\nu < 0.312$,
while negative values have $-D N_\nu < 0.350$.
We see that the difference is too
small to accommodate a full scalar appearing or
disappearing between the two epochs.

\begin{figure}
    \centering
    \includegraphics[width=\textwidth]{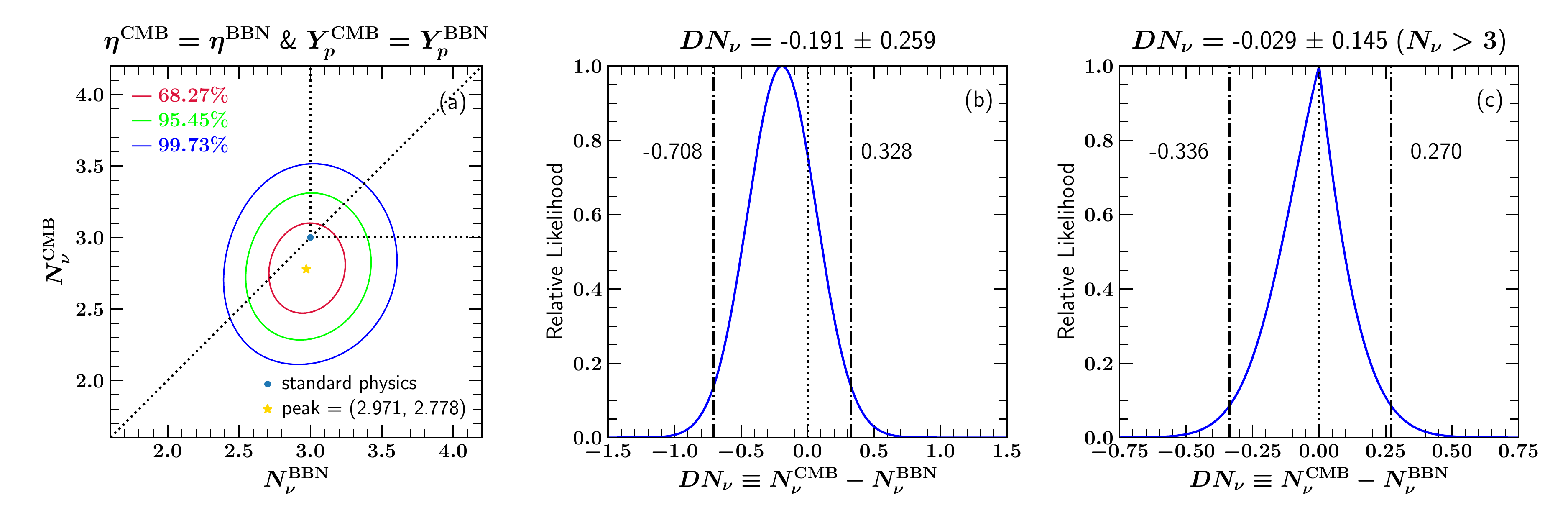}
    \caption{CMB vs BBN determinations of $N_\nu$. Similar to Fig. \ref{fig:Nnu_compare_same_Yp} but additionally demanding $\eta^{\rm CMB} = \eta^{\rm BBN}$. 
    }
    \label{fig:Nnu_compare_same_EtaYp}
\end{figure}

In principle, there could be scenarios where $N_\nu$ changes between  BBN and the CMB but $\eta$ does not. Therefore 
it is appropriate to consider a likelihood function constrained by $\eta^{\rm BBN} = \eta^{\rm CMB}$ which gives
\beq
\label{eq:same-eta}
{\cal L}(N_\nu^{\rm BBN},N_\nu^{\rm CMB})
\ \propto \ \int {\cal L}(N_\nu^{\rm BBN}, \eta, \ N_\nu^{\rm CMB}, \eta) 
\ ; \ \  d\eta \qquad \eta = \eta^{\rm BBN} = \eta^{\rm CMB}
\eeq
where the integrand is given by Eq.~(\ref{eq:4D}).
This likelihood distribution is expected to give stronger limits on the change in $N_\nu$ if
the BBN- and CMB-preferred $\eta$ values are similar.  
Results for this case appear in Fig.~\ref{fig:Nnu_compare_same_EtaYp}.  
We see that the results for each panel
are quite similar to those from the corresponding
part of Fig.~\ref{fig:Nnu_compare_same_Yp},
but as expected the limits are tighter.
Again all three panels show results fully consistent
with the Standard Model.
In panel (b), the $\Delta N_\nu$ is slightly
shifted to negative values, but by less than $1\sigma$.

These limits place new constraints on many kinds of models for physics in the early universe; we now discuss a few examples.

\paragraph{Early Dark Energy models.}  A change in $N_\nu$ between BBN and the CMB occurs in Early Dark Energy models \cite{Poulin2019}.
For the case of an oscillating scalar field in a potential which is approximately quartic about the minimum, the oscillations act like radiation with an equation of state $w=1/3$.  Using best-fit values, the dark energy component acts
as radiation for scale factors $a>a_{\rm c} \approx 1.9 \times 10^{-4}$ or $z_{\rm c} \approx 5300$,
close to the epoch of matter-radiation equality
when $a_{\rm eq} = \Omega_{\rm r,0}/\Omega_{\rm m,0} \approx 1/3400$.
Thus, during recombination this component
will act as radiation, adding to $N_\nu$.\footnote{In the context of self-interacting coherent scalar dark matter models, similar constructions can instead lead to decreases in $N_\nu$ between BBN and the CMB \cite{Li:2013nal}.}
At $z_c$, the model requires a best-fit $f_{\rm EDE} = \Omega_\phi(a_c) = 0.044$,
assuming $\Omega_{\rm tot}=1$.
Compared to the ordinary radiation density, we have $\rho_\phi(a_c)/\rho_r(a_c) = f_{\rm EDE} (1 + a_c/a_{\rm eq})$. We can write $\rho_\phi/\rho_r \approx g_{1\nu} \, D N_\nu^{\rm EDE}/(2 + 3 g_{1\nu})$, with $g_{1\nu} = 7/4 \ (4/11)^{4/3}$.  We thus 
cast the early dark energy perturbation during the CMB as
\beq
D N_\nu^{\rm EDE} \approx  f_{\rm EDE} \left( 1 + \frac{a_c}{a_{\rm eq}} \right) 
\left[ 3 + \frac{8}{7} \pfrac{11}{4}^{4/3} \right] 
= 0.53 \ \ ,
\label{EDE}
\eeq
an increase of about half of an equivalent neutrino species at recombination.  
Comparing to our results from this section,
we see that this is allowed when $N_\nu$ is free, but is not allowed (at 2 $\sigma$) 
when $N_\nu > 3$ in
Fig.~\ref{fig:Nnu_compare_same_Yp}.
Furthermore, from Fig.~\ref{fig:Nnu_compare_same_EtaYp},
we see that when $\eta$ is held fixed between BBN and CMB decoupling, $DN_\nu < 0.33$ putting still more pressure on this model. Indeed as we will see in the next subsection, to attain $DN_\nu^{\rm EDE}$
as large as that in Eq.~(\ref{EDE}), would require an increase in $\eta$ between the BBN and CMB epochs.
This model--or more precisely, this choice of $(f_{\rm EDE},a_c,w)$--is thus at best marginally allowed from our viewpoint.  Moreover, this
example shows that there is a role for
constraints of the type we have presented.
Other early dark energy models with different potentials and hence different equations of state  parameters $w$ will
also be constrained by BBN + CMB limits, but these cannot be parameterized by $N_\nu$ 
and so would require dedicated study.  

\paragraph{Relativistic relic becoming nonrelativistic.} 
A BSM particle $\chi$ that contributes to $N_\nu$ during BBN may become nonrelativistic prior to the formation of the CMB.  In this case the number of effective relativistic degrees of freedom will decrease between the two epochs, leaving $\eta$ unaffected.
The contribution of such a free-streaming relativistic relic to $\Neff$ during BBN can be written in a very general way as 
\beq
\rho_\chi|_{BBN}  = \left(\frac{a_{\mathrm{nr}}}{a_{\mathrm{eq}}}\right) \fdm  \times \left(\frac{\Omega_{\rm cdm}}{\Omega_{\rm m}}\right) \rho_{SM}|_{BBN} ,
\eeq
where  
$a_{\mathrm{nr}}$ is the scale factor when the relic becomes nonrelativistic and
$a_{\mathrm{eq}}$ is the scale factor at matter-radiation equality, with $a_{\mathrm{nr}} < a_{\mathrm{eq}}$, $\fdm    <1$ is the fraction of dark matter  contributed by the relic after it becomes nonrelativistic, and $\Omega_{\rm cdm},\, \Omega_{\rm m}$ are the fractions of critical density in cold dark matter and total matter (CDM $+$ baryons). 
Here we have assumed only 
that the evolution of the energy density of $\chi$ can be well-approximated as redshifting like radiation for $a< a_{\mathrm{nr}}$ and matter for $a>a_{\mathrm{nr}}$.\footnote{This is a reasonable approximation for relics whose momentum distribution is dominated by a single scale, such as warm dark matter or frozen-in dark matter, but may break down for relics that have multiple features in their phase space distribution.}
Specific models of warm dark matter, such as sterile neutrinos, will relate $\fdm    $ and $a_{\mathrm{nr}}$ to the mass and production mechanism of the hot relic $\chi$.
Given this expression for the energy density for the relativistic relic,  the corresponding shift in  $N_\nu$ is
\[
- D N_\nu  
=  6.2  \left(\frac{a_{\mathrm{nr}}}{a_{\mathrm{eq}}}\right) \fdm     ,
\]
where we have used the results of Ref.~\cite{Planck:2018vyg} for cosmic parameters.
Imposing the one-sided constraint $-DN_\nu < 0.336$, we thus obtain
\beq
\label{eq:maxnr}
\left(\frac{a_{\mathrm{nr}}}{a_{\mathrm{eq}}}\right) \fdm     < 0.054.
\eeq
For fractions $\fdm    $ near unity, this result constrains relics becoming nonrelativistic at sufficiently early times that the assumption of a constant value of $N_\nu$ at the CMB epoch is consistent. 

This dark radiation disappearance limit depends on two key properties of the relic: (i) the scale $a_{\mathrm{nr}}$ at which it becomes nonrelativistic, and (ii) the fraction $\fdm    $ of dark matter that the relic represents after it becomes nonrelativistic.  These two properties are also exactly the same  quantities that are important for understanding the potential impact of the hot relic on structure formation, which lets us make some general observations about $DN_\nu$ limits compared to those arising from measurements of the Lyman-$\alpha$ forest.  The free-streaming horizon
\beq
\lambda_{FS} =  \int_{t_0}^{t_f} \frac{v}{a} dt \approx \frac{1}{a_{\mathrm{nr}} H_{nr}} \left[  1+\ln \left(a_{\mathrm{eq}}/a_{\mathrm{nr}}\right) \right] 
\eeq
is a useful estimate for the maximum scale on which a hot relic can suppress the growth of density perturbations in the early universe.  While Lyman-$\alpha$ constraints on warm relics require specifying the model-dependent phase space distribution of the relic, when $\fdm    =1$ (and again assuming a phase-space distribution characterized by a single momentum scale), it is possible to extract a conservative and model-insensitive requirement that $\lambda_{FS}\lesssim 0.1$ Mpc in order to match current observations \cite{Ballesteros:2020adh}. 
Approximating the Hubble rate as piecewise power laws in the scale factor $a$ lets us write 
\beq
\label{eq:fs}
\lambda_{FS} \approx \frac{1}{H_0}  \sqrt{\frac{\Omega_R \Omega_\Lambda}{\Omega_M^2}} \frac{a_{\mathrm{nr}}}{a_{\mathrm{eq}}} \left[  1+\ln \left(a_{\mathrm{eq}}/a_{\mathrm{nr}}\right) \right].
\eeq
For $\fdm    =1$ the $DN_\nu$ limit on $a_{\mathrm{nr}}/a_{\mathrm{eq}}$ in eq.~(\ref{eq:maxnr}) translates into the weaker constraint $\lambda_{FS} \lesssim 24$ Mpc using eq.~(\ref{eq:fs}). Thus for models of light relics, independent of the detailed microphysics in a given model, the present $DN_\nu$ limit is meaningful but less constraining than the current Lyman-$\alpha$ limits (for more model-specific statements see \cite{Ballesteros:2020adh,Das:2020nwc,Li:2021okx,Decant:2021mhj}). 


\paragraph{Late equilibration with neutrinos.}  BSM physics that has relevant interactions with the SM neutrinos can come into equilibrium with the SM neutrinos after BBN and subsequently become nonrelativistic, depositing their entropy into the neutrino bath \cite{Berlin:2017ftj,Berlin:2018ztp,Berlin:2019pbq,Kelly:2020aks}.  In such scenarios the effective number of neutrinos increases between BBN and the CMB, while leaving $\eta$ and $Y_p$ unaffected.
The one-sided constraint $DN_\nu < 0.270$ 
allows for a  single relativistic degree of freedom to equilibrate with the SM neutrinos provided its temperature before BBN was no more than $0.6 T_\gamma$. 
Late equilibration with a particle species with more than one degree of freedom cannot be accommodated.

\paragraph{Inflationary dark vectors decaying to neutrinos.}
Dark vector bosons produced through inflationary fluctuations can decay to SM neutrinos if they couple to the $B-L$ current, or similarly to the current for one of the other anomaly-free but non-flavor-universal global $U(1)$ symmetries of the SM, $L_i - L_j$ or $B-3 L_i$ \cite{Krnjaic:2020znf}.  When this decay occurs between neutrino decoupling and recombination, it gives rise to a shift in $N_\nu$ while leaving $\eta$ unaffected.  The vector boson's mass controls the timing of its decay, but not the size of the resulting shift in $N_\nu$; instead, this shift depends quadratically on the (unknown) Hubble scale during inflation and inversely on the vector boson's coupling to neutrinos \cite{Krnjaic:2020znf}. 
The constraints on this scenario from the limit $DN_\nu < 0.270$ are slightly more stringent than those shown in \cite{Krnjaic:2020znf}, which used a CMB+LSS $\Delta N_{\mathrm{eff}}$ constraint of $\Delta N_{\mathrm{eff}} < 0.28$.

\subsection{Limits on the Evolution of the Baryon-to-Photon Ratio}

In this section, we place limits on 
changes in the baryon-to-photon ratio
between BBN and CMB decoupling.
As shown in Eq.~(\ref{eq:eta-entropy}),
changes in comoving entropy lead to changes in $\eta$.
In the standard picture $e^\pm$ pair annihilation transfers entropy to photons during BBN,
and BBN calculations include these effects.  Here we look for changes in
$\eta$ beyond this, and between the two epochs.

Nonstandard scenarios with entropy change include particle creation or destruction,
e.g., by out-of-equilibrium decays.
As noted in the discussion surrounding
Eq.~(\ref{eq:eta-entropy}), entropy change typically also
changes $N_\nu$, and such cases are treated
in the next subsection.
Indeed, it is challenging to try
to construct physically motivated
scenarios that change $\eta$
while holding $N_\nu$ fixed.
Nevertheless, for completeness we show results
for such scenarios here.

\begin{figure}
    \centering
    \includegraphics[width=\textwidth]{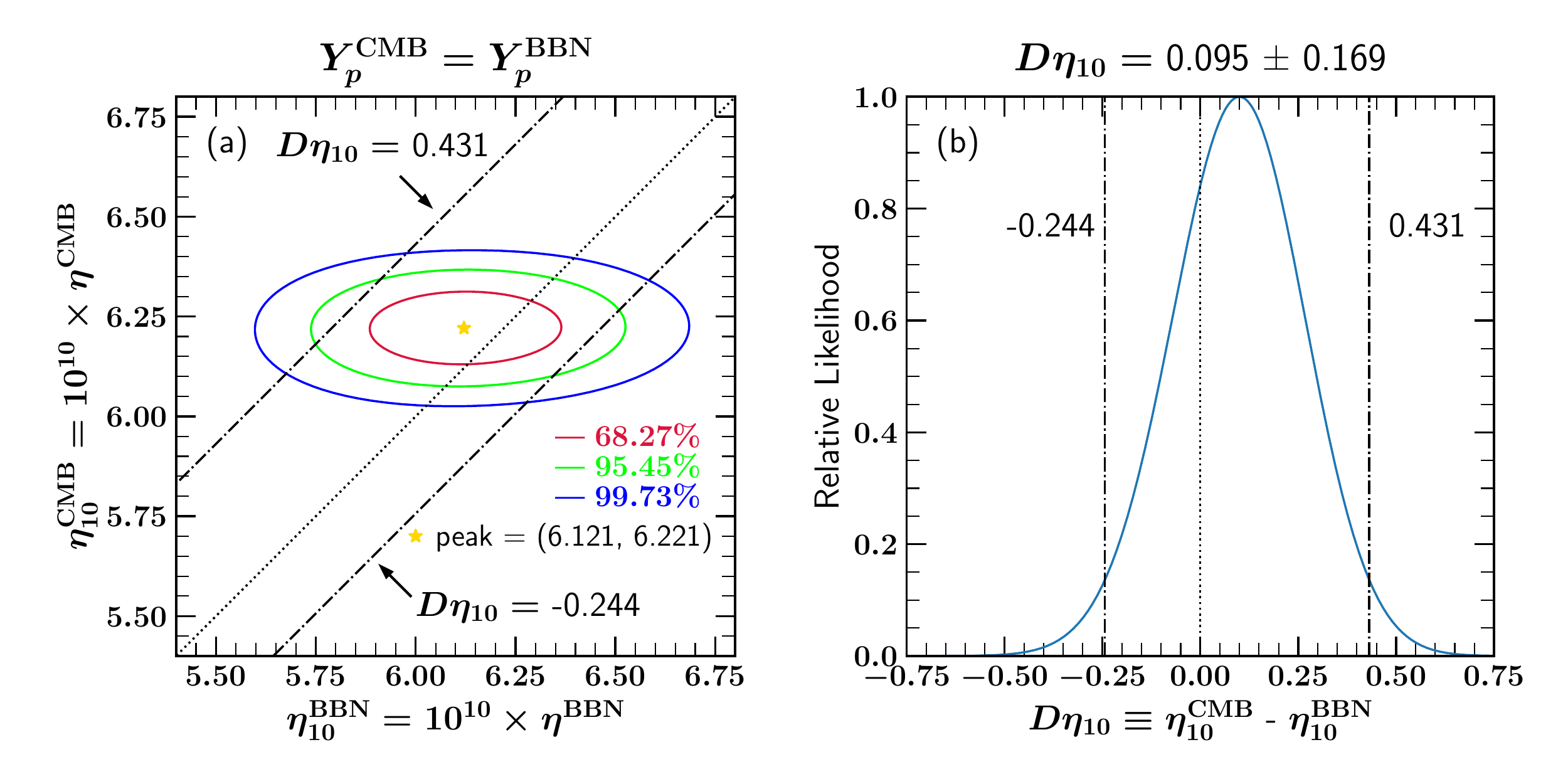}
\caption{A comparison of the independent BBN and CMB measures of $10^{10} \eta$, marginalizing separately over all $N_\nu$ values but holding $Y_p$ fixed as in Fig.~\ref{fig:Nnu_compare_same_Yp}.
Panel (a) shows $\eta^{\rm CMB}$ vs $\eta^{\rm BBN}$ likelihood contours.  We see that the best-fit point lies within $1\sigma$ of the
Standard Cosmology $\eta^{\rm CMB}=\eta^{\rm BBN}$ shown by the dotted line.
We also see that $\eta^{\rm CMB}$ constraints are significantly tighter.
Panel (b) shows the distribution for the {\em difference}
$D \eta = \eta^{\rm CMB}-\eta^{\rm BBN}$.  
The dot-dashed lines show the the two-sided 95\% CL limits, and
these also appear as the dot-dashed diagonal lines in (a).
We see the peak lies near the Standard Cosmology $D \eta = 0$ value, and that there are rather tight limits on deviations away from this value.
} 
    \label{fig:eta_compare_same_Yp}
\end{figure}

For the most conservative limits, we marginalize
separately over $N_\nu^{\rm BBN}$ and
$N_\nu^{\rm CMB}$:
\begin{equation}
    {\cal L}(\eta^{\rm BBN}, \eta^{\rm CMB}) = \int {\cal L}(N_\nu^{\rm BBN}, \eta^{\rm BBN}, N_\nu^{\rm CMB},\eta^{\rm CMB}) \ dN_\nu^{\rm BBN} \ dN_\nu^{\rm CMB}  \ \ .
\end{equation}
Note that the 4-D integrand imposes $Y_p^{\rm CMB}=Y_p^{\rm BBN}$
via Eq.~(\ref{eq:4D}).
Results appear in Fig.~\ref{fig:eta_compare_same_Yp}.
We see that the peak values of $\eta^{\rm CMB}$ and $\eta^{\rm BBN}$ are indeed quite close to each other, less than $1\sigma$ apart.  They thus are close to dotted diagonal line giving the Standard Model case $\eta^{\rm CMB} = \eta^{\rm BBN}$.  This is a manifestation of the BBN-CMB concordance that stands as a success of the hot big bang theory.
The plot also makes clear the well-known result that the CMB measurement of $\eta$ is substantially tighter than that from BBN.
Finally, we see that there is not a significant correlation between
the two $\eta$ values, despite using a common $Y_p$ constraint.  This is because $Y_p$ does not play an important role in setting $\eta^{\rm BBN}$. 

We denote the change in baryon-to-photon ratio as
\begin{equation}
    D\eta = \eta^{\rm CMB} - \eta^{\rm BBN} \ \ ,
\end{equation}
and for convenience we also use $D\eta_{10} = 10^{10} D \eta$.
Panel (b) of Fig.~\ref{fig:eta_compare_same_Yp}
shows the distribution of this difference.
Each value of $D \eta_{10}$
corresponds to a diagonal in panel (a), and
we marginalize over this diagonal to find the value
at each point in panel (b).  
We see that $D \eta_{10}$ peaks close to
the Standard Model $D\eta = 0$ value,
with a slight preference for positive values.
As a result, the limits on negative $D\eta$
are stronger than those for positive values.

We saw in Eq.~(\ref{eq:eta-entropy}) that $\eta \propto g_{*S}/S_{\rm com}$:  an increase in the entropy of the baryon-photon plasma leads to a decrease in $\eta$.  
With this in mind, we note that the 1-sided limit on 
a decrease
$D\eta_{10} < 0$, 
\beq
\left. D\eta_{10} \right|_{\rm min} = -0.287  
\eeq
at $2\sigma$ (the 2-sided limits are shown in Fig.~\ref{fig:eta_compare_same_Yp}).  Given the mean $\eta_{10} \sim 6$,
this means only a $\sim 5\%$ decrease is allowed.

\begin{figure}
    \centering
    \includegraphics[width=\textwidth]{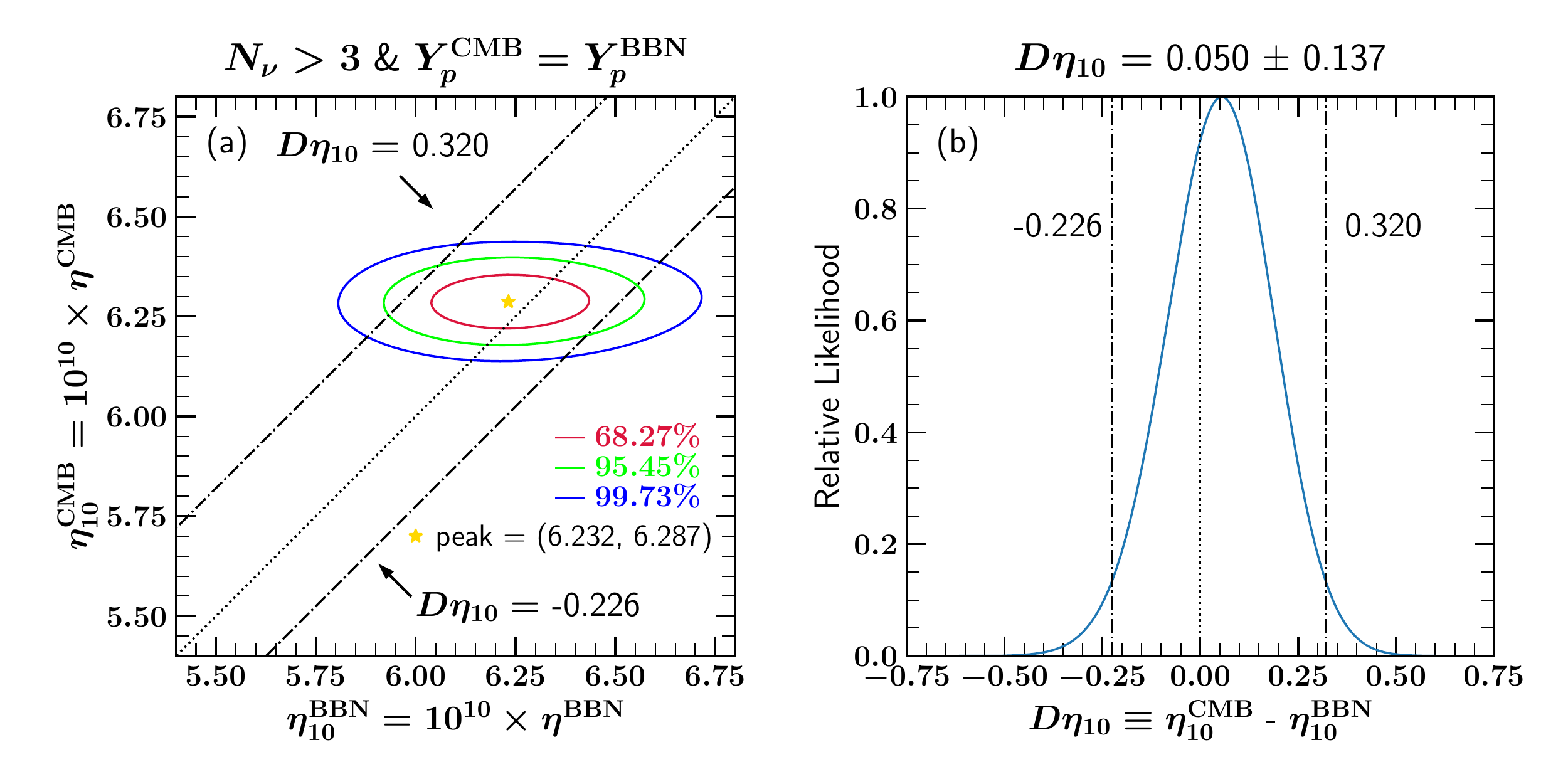}
\caption{Similar to Fig.~\ref{fig:eta_compare_same_Yp}, but requiring $N > 3$ for both the CMB and BBN.
} 
    \label{fig:eta_compare_same_Yp_Ngt3}
\end{figure}

We have also examined possible changes in $\eta$ in the
case where we require $N_\nu >3$
at both BBN and CMB epochs.
These results appear in Fig.~\ref{fig:eta_compare_same_Yp_Ngt3},
where the main lesson is that the consistency with the Standard Model is even better,
with the peaks in both panels even closer to 
$D\eta=0$.
The one-sided limit on the decrease 
$D\eta_{10} < 0$ now becomes
\beq
\left. D\eta_{10} \right|_{\rm min,N>3} = -0.251  
\eeq
at $2\sigma$ (the 2-sided limits are shown in Fig.~\ref{fig:eta_compare_same_Yp_Ngt3}), now allowing only a $\sim 4\%$ difference.
We see that very little change is allowed in the baryon-to-photon ratio, and thus in entropy, between
BBN and the CMB.

\subsection{Limits on Evolution of Both $\eta$ and $N_\nu$}

We now relax the assumption that only one 
of $\eta$ and $N_\nu$ varies, and we allow for 
both to change between nucleosynthesis and recombination.
Examples where this situation can occur 
would be entropy-producing scenarios such as annihilations or decays into photons. 

For our most conservative case, 
we do not require $Y_p^{\rm CMB}=Y_p^{\rm BBN}$. 
Rather, we use equation (\ref{eq:LBBN}) for ${\cal L}_{\rm NBBN+obs}(\eta, N_\nu)$, and 
marginalize over $Y_p$ to find
\begin{equation}
{\mathcal L}_{\rm NCMB}(\eta, N_\nu) \ \propto \  \int  {\mathcal L}_{\rm NCMB}(\eta,N_\nu,Y_p) \ dY_p \, .
\end{equation}
We then calculate the likelihoods
for the differences $D \eta = \eta^{\rm CMB}-\eta^{\rm BBN}$
and $DN_\nu = N_\nu^{\rm CMB}-N_\nu^{\rm BBN}$:
\begin{equation}
{\mathcal L}_{\rm NCMB+NBBN+obs}(D \eta, D N_\nu) \ \propto \  \int  {\mathcal L}_{\rm NCMB}(\eta +D \eta, N_\nu + D N_\nu){\cal L}_{\rm NBBN+obs}(\eta, N_\nu) \ d\eta \, dN_\nu \ .
\end{equation}
This convolution takes the 4D distribution in Eq.~(\ref{eq:4D}) down to a 2D distribution by focusing on the differences in $\eta$ and $N_\nu$.

If we do require that
the BBN and CMB $Y_p$ values are the same,
as in Eq.~(\ref{eq:He-equality}),
then the convolution
\begin{equation}
    {\cal L}_{\rm NBBN+NCMB+obs}(D \eta, D N_\nu)
    \ \propto \ 
    \int {\cal L}_{\rm NCMB}(\eta +D \eta, N_\nu + D N_\nu, Y_p) \ 
    {\cal L}_{\rm NBBN}(\vec{X};\eta, N_\nu) \ 
    \prod  {\cal L}_{\rm obs}(X_i) \ dX_i \ d\eta \ dN_\nu \ 
\end{equation}
gives the likelihood distribution for the change in these parameters.

\begin{figure}
    \centering
    \includegraphics[width=\textwidth]{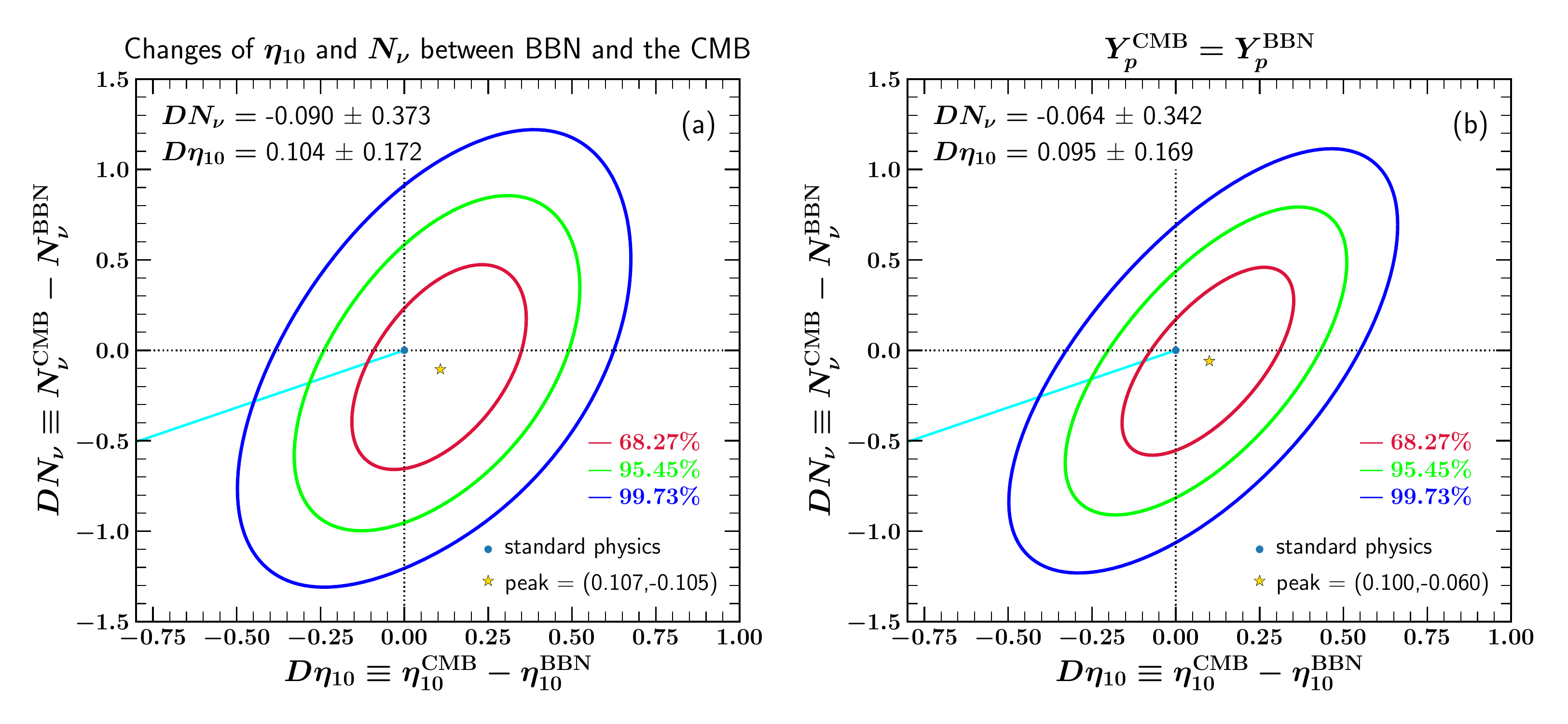}
    \caption{Allowed variation in both $\eta$ and $N_\nu$ between BBN and the CMB, 
    assuming $Y_p$ is the same in both epochs in panel (b).  The cyan lines are for the scenario described in eqs.~(\ref{eq:Deta}-\ref{eq:DNnu}), for the case of $f_\gamma =1$.
     }
    \label{fig:NEchange}
\end{figure}

\begin{figure}
    \centering
    \includegraphics[width=\textwidth]{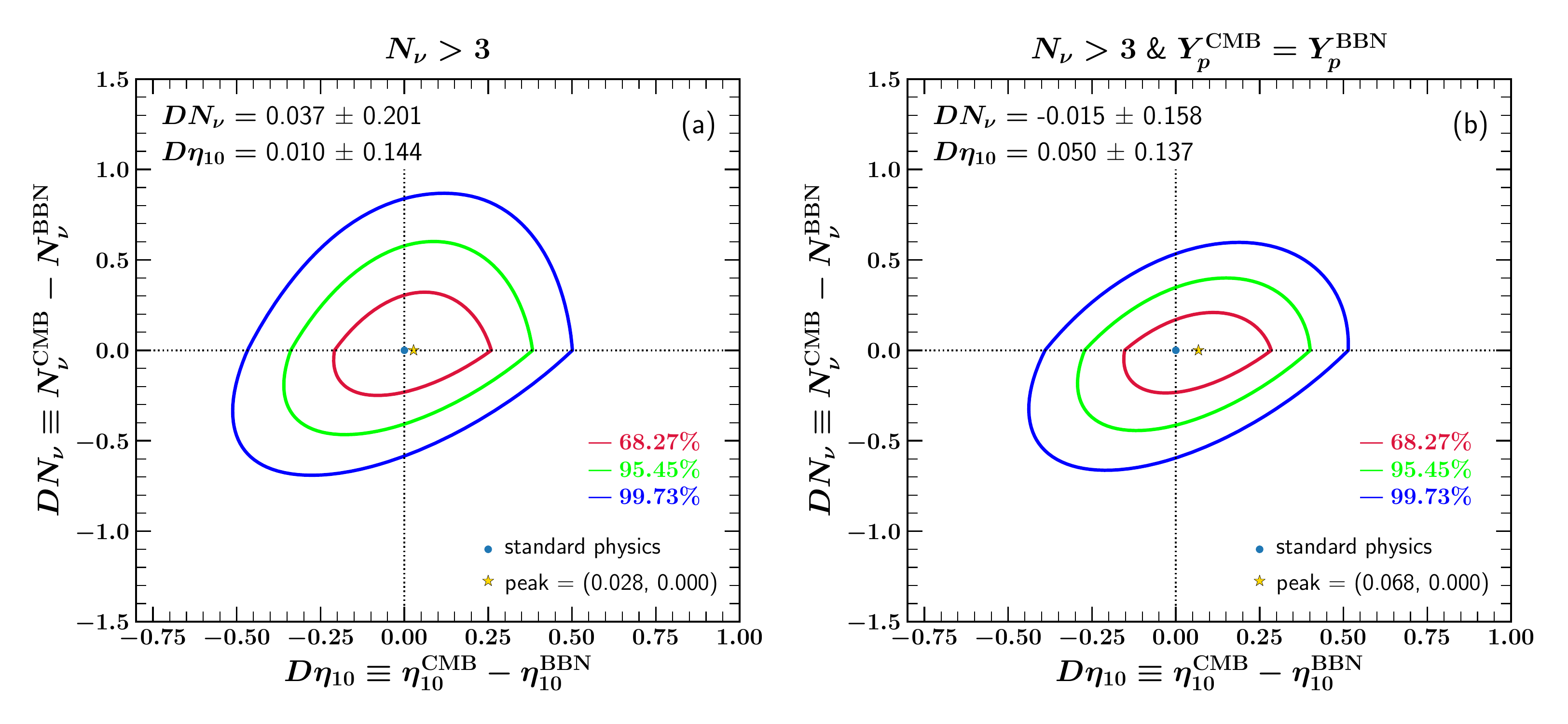}
    \caption{Similar to Fig.~\ref{fig:NEchange}, but now requiring $N > 3$.
     }
    \label{fig:NEchange_Ngt3}
\end{figure}

Fig.~\ref{fig:NEchange} shows the allowed
changes in $\eta$ and $N_\nu$ for both
cases.
Remarkably, but at this point not surprisingly in panel (a) and (b) the maximum likelihood is close to $(0,0)$. That is, even with this greater freedom, we again see that the Standard Model gives an excellent fit, and the data do not show a need for variation between the two epochs.
We also see a positive correlation between
$D\eta$ and $DN_\nu$.  This reflects the fact 
shown in Fig.~\ref{fig:Nnu_vs_eta} that
BBN alone (and to lesser extent, the CMB alone)
has a positive $(\eta,N_\nu)$ correlation.
Figure~\ref{fig:NEchange_Ngt3} is similar to
Fig.~\ref{fig:NEchange},
but now allowing only $N_\nu >3$ for both BBN and the CMB.  
We see in Fig.~\ref{fig:NEchange}
that the regions with $DN_\nu < 3$ allow for larger deviations between the two epochs.  
Thus when we exclude these, the result is mostly a narrowing of the allowed $N_\nu$ range.

Figures~\ref{fig:NEchange} and \ref{fig:NEchange_Ngt3} also generalize
some of our earlier results.
The $N_\nu$ change for $\eta^{\rm CMB}=\eta^{\rm BBN}$
corresponds to a vertical line at $D\eta=0$;
this is what is shown in Fig.~\ref{fig:Nnu_compare_same_EtaYp}.

One immediate application of these constraints is a species that decays out of equilibrium in the epoch  between BBN and recombination, which  can inject sufficient energy and entropy into the SM radiation bath to alter both $\Neff$ and $\eta$.  
Related work has focused on particle decays
\cite{Scherrer:1984fd,Fischler:2010xz,Menestrina:2011mz,DiBari:2013dna}.
We illustrate the impact of this varying-$\eta$, varying-$N_\nu$ analysis with a general example where a decaying particle injects energy density $\Delta \varepsilon$ into the photon bath after the conclusion of BBN and before the formation of the CMB. We consider the case where all of this energy is deposited into photons at redshifts $z\gtrsim \mathrm{few}\,\times 10^6$, so that it can be simply parameterized by a shift in the photon temperature \cite{Chluba:2013kua}.  In this redshift range constraints on $DN_\nu$ and $D\eta$ provide a leading probe of energy injection in the early universe, while spectral distortions become more powerful at lower redshifts \cite{DeZotti:2020glv}.\footnote{We restrict attention to the case where the energy density of the decaying species is negligible during BBN; see Ref.~\cite{Sobotka:2022vrr} for a more detailed analysis where this assumption is relaxed.}

To leading order in the fractional energy injection $\Delta \varepsilon/\rho_\gamma^{CMB}$, the shift in $\eta$ and $N_\nu$ resulting from such an energy injection can be written
\begin{eqnarray}
\label{eq:Deta}
D\eta & \approx &
 \eta^{CMB} \left[-\frac{3}{4} \frac{\Delta\varepsilon}{\rho_\gamma^{CMB}}\right] \\
 \label{eq:DNnu}
DN_\nu &\approx & 
 N_\nu^{BBN} \left[ -\frac{\Delta\varepsilon}{\rho_\gamma^{CMB}} \right],
\end{eqnarray}
where $\Delta\varepsilon$ parameterizes the net energy deposited into photons after the conclusion of BBN, such that $\rho_\gamma^{CMB} = \rho_\gamma^{BBN} (a_{BBN}/a_{CMB})^4 + \Delta\varepsilon $. 

We can now constrain the fractional energy release $\Delta\varepsilon/\rho_\gamma^{\rm CMB}$ 
using the 2-D joint constraints from from Fig.~\ref{fig:NEchange}.
As $\Delta\varepsilon/\rho_\gamma^{\rm CMB}$ 
varies from $0$ and up, the constraints
in eqs.~(\ref{eq:Deta}-\ref{eq:DNnu})
describe a line in the $(D\eta,DN_\nu)$ plane,
taking $N_{\nu}^{\rm BBN}$ and $\eta^{\rm CMB}$ as fixed.
This line appears in 
Fig.~\ref{fig:NEchange}(a),
and intersects the $2\sigma$ contour
at $(D\eta,DN_\nu) = (-0.28, -0.18)$. 
These correspond respectively to limits 
$\Delta \varepsilon/\rho_\gamma^{\rm CMB}
< (0.061, 0.062)$ which are almost equal,
and so the $D\eta$ constraint drives the limit
\begin{equation}
\label{eq:Einj}
\frac{\Delta \varepsilon}{\rho_\gamma^{\rm CMB}}
< 0.061 \ \ .
\end{equation}
Here we adopted the central value for $\eta_{\rm CMB}$,
but the uncertainty on this quantity is $< 1\%$, and so our limit is essentially unchanged.
Note that in this case, the analysis shown
in Fig.~\ref{fig:NEchange_Ngt3} does not apply,
because it assumes $N_\nu \ge 3$ for both the BBN
and CMB epochs.  But even in the presence of SM neutrinos, in this example the heating of photons
relative to neutrinos will lead to $T_\nu/T_\gamma < (4/11)^{1/3}$, which would show up in CMB analyses
as $N_\nu^{\rm CMB}< 3$.

In a case where the decaying species deposits a fraction $\fgam$ of its energy into photons while the remaining fraction $1-\fgam$ is deposited into neutrinos, the resulting shifts in $\eta$ and $N_\nu$ become, again to leading order in  $\Delta \varepsilon/\rho_\gamma^{CMB}$,
\begin{eqnarray}
\label{eq:DNnu_f}
DN_\nu & \approx& N_\nu^{BBN} \left[  \frac{1-\fgam}{c_\nu N_\nu^{BBN}} -\fgam  \right] \frac{\Delta \varepsilon}{\rho_\gamma^{CMB}} \\
D\eta &\approx &  \eta^{CMB} \fgam \left[-\frac{3}{4} \frac{\Delta\varepsilon}{\rho_\gamma^{CMB}}\right],
\end{eqnarray}
where $c_\nu = \frac{7}{8} \left(\frac{4}{11}\right)^{4/3}$.
Using $N_\nu^{\rm BBN}=3$,
the bracketed coefficient in eq.~(\ref{eq:DNnu_f})
goes from $-1$ to $+1.47$ as $\fgam$ decreases from
1 to 0.  
There is a finely-tuned value of $\fgamc = (1+N_{\nu}^{\rm BBN}c_\nu)^{-1} = 0.595$ for which the energy of the decaying particle is shared between photons and neutrinos in a ratio that exactly replicates that predicted by the standard cosmology, resulting in $DN_\nu = 0$, while $\eta$ still decreases.
Thus, the $DN_\nu$ variation is largest and positive for $\fgam=1$, vanishes for $\fgam=\fgamc$
and reverses sign for $\fgam<\fgamc$.
Also, for all $\fgam$
the $DN_\nu/N_\nu-D\eta/\eta$ correlation is still
linear, but the slope is generally shallower than for $\fgam=1$, becoming negative for $\fgam < \fgamc$.
For $\fgam = 0.5$, we find that the $D\eta$ constraint
in Fig.~\ref{fig:NEchange}(a) is stronger, and 
with $D\eta/\eta^{\rm CMB} < -0.36$ 
gives
$\Delta \varepsilon/\rho_\gamma^{\rm CMB} < 0.096$, almost a factor of 2 weaker than the result in eq.~(\ref{eq:Einj}).

The photon bath can also experience a net entropy increase if it comes into equilibrium with a feebly-coupled dark sector after BBN \cite{Berlin:2017ftj,Berlin:2019pbq}.
This late equilibration can be naturally realized in theories where a light (sub-MeV) species mediates interactions between photons and one or more SM singlet fields. 
Concrete models that realize late equilibration with the photon plasma are however subject to a number of stringent constraints, particularly from stellar cooling \cite{Berlin:2019pbq}.

\section{The Expected Impact of Future Observations: CMB-S4, Precision $^4$He Observations and the Hunt for Neutrino Heating}
\label{sect:future}

We look forward to a bright future in which the constraints we have presented will become stronger.
For the microwave background,
CMB Stage-4 (CMB-S4) will be the next generation ground-based CMB experiment. CMB-S4 is expected to improve the CMB precision at small angular scales, reducing uncertainties at high multipoles in CMB anisotropy power spectra. Such improvement will result in a better CMB measurement of $\sigma(N_{\nu})$.
On the BBN side, the $Y_p$ uncertainties from astronomical observations currently dominate the size of BBN $\sigma(N_{\nu})$. Therefore, we also look forward to future improvements on the precision of primordial helium observations to provide BBN-only $N_{\nu}$ determination that can compete with the expected CMB-S4 results. 

\begin{figure}
    \centering
    \includegraphics[width=\textwidth]{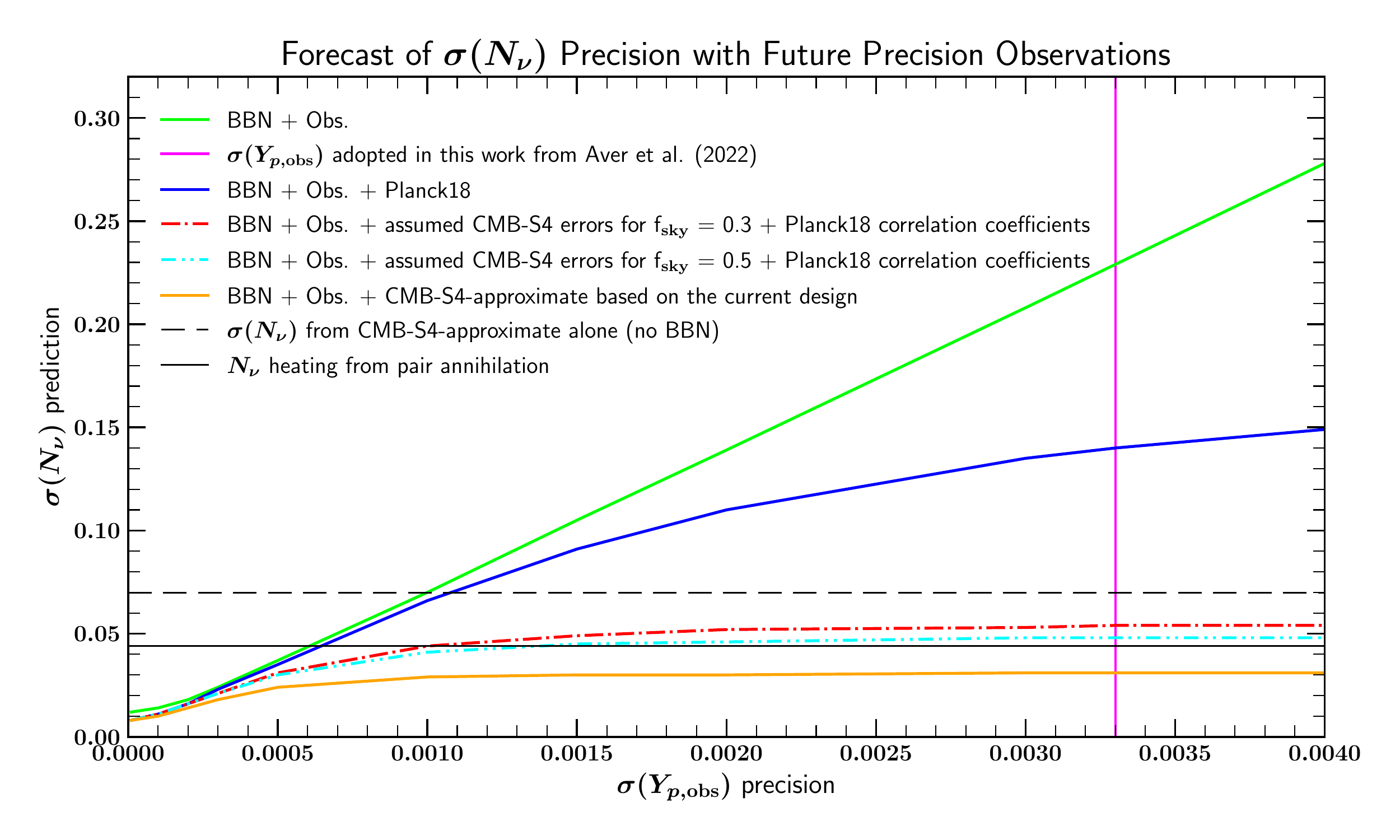}
    \caption{Impact on the precision of $N_\nu$ due to future improvements in astronomical \he4 measurement precision. Curves show the $\sigma(Y_{p,\rm obs})$ impact on 
    $\sigma(N_\nu)$ for the cases of (a) BBN+Obs., which in green shows a nearly linear scaling down for most of the range, and of (b) BBN+Obs.+CMB, where the shallower and nonlinear responses reflect the contribution of past and future CMB constraints on $N_\nu$ (in blue and orange respectively). The vertical line for the current $\sigma(Y_{p,\rm obs})$, the horizontal dashed line for the expected $\sigma(N_\nu)$ from the CMB-S4-approximate case alone, and the horizontal solid line for the impact of neutrino heating are three auxiliary lines for reference. We see it is very promising to probe the neutrino heating effect by combining current BBN constraints with the future CMB-S4 data.
    }
    \label{fig:futureYp}
\end{figure}

Figure \ref{fig:futureYp} shows
forecasts for the uncertainty
in $N_\nu$
in response to improvements in astronomical
measurements of $Y_p$.
The different curves show the
expected $\sigma(N_\nu)$
given the observed $\sigma(Y_p)$
error budget, for different combinations
of BBN with CMB measurements past and future.
For BBN alone (shown in green), we see that the trend is linear
over most of the domain; this reflects
the fact that for BBN only, $Y_p$
dominates the $N_\nu$ inference.\footnote{
The slope of this trend fits well the 
expectations from the scaling
$Y_p \propto N_\nu^{0.163}$
that gives $\sigma(N_\nu) = (1/0.163)(N_\nu/Y_p) \, \sigma(Y_p) \approx 70 \, \sigma(Y_p)$.
}
We see that for BBN alone,
to begin to see the effects of neutrino heating
(Eq. \ref{eq:nuheat})
with $\sigma(N_\nu) < 0.044$ (shown by the solid-black line) requires
very precise helium determinations:
$\sigma(Y_p) < 0.0006$.

Figure \ref{fig:futureYp} also highlights the dramatic improvements when BBN and CMB measurements are combined (the solid-blue curve).  We see that adding the {\em Planck} 2018 
information dramatically improves
the $N_\nu$ sensitivity for $Y_p$
measurements at or somewhat below the current levels. At the current $\sigma(Y_p) = 0.0033$,
we see that BBN+CMB combination improves
the $\sigma(N_\nu)$ by about a factor of 2.
Note that the BBN-only and CMB-only errors
are comparable, so the effect is not just
one of averaging, but rather the combination
breaks degeneracies and so is quite powerful.
There remains significant improvement
down to $\sigma(Y_p) \simeq 0.0012$, but
to reach the neutrino heating still
requires similar \he4 precision to the BBN-only case.  

Looking forward to CMB-S4, we find
the BBN+CMB-S4 combination should be very powerful.
The CMB-S4 constraints are sensitive to the fraction $f_{\rm sky}$ of the sky observed.
We will use $f_{\rm sky}^{\rm S4} = 0.3$ as a baseline for CMB-S4, but
a larger sky coverage would tighten the limit.
We assume in our toy model that the CMB-S4 likelihood is a multivariate Gaussian distribution and has $\sigma(N_{\rm eff}^{\rm S4}) = (0.09, 0.08)$\footnote{When referring to the CMB neutrino determination, we use $N_{\rm eff}$ to follow the convention in the CMB literature. See Eq. (\ref{NeffNnu}) for the $N_{\rm eff}-N_\nu$ relation.} and $\sigma(Y_p^{\rm S4}) = (0.0055, 0.0047)$ for $f_{\rm sky}^{\rm S4} = (0.3, 0.5)$.\footnote{We also assume $\sigma(\eta^{\rm S4}) = 0.02\times 10^{-10}$ for both cases, which is about a factor of 3 improvement from {\it Planck} 2018 similar to the assumed improvements of $\sigma(N_{\rm eff}^{\rm S4})$ and $\sigma(Y_p^{\rm S4})$ in our toy model.} These values are approximately inferred from the Figure 28 of the CMB-S4 science book \cite{CMB-S4:2016ple}, which provides forecasts for $\sigma(N_{\rm eff}^{\rm S4})$ and $\sigma(Y_p^{\rm S4})$ as functions of sky fraction.\footnote{Both of $N_{\rm eff}^{\rm S4}$ and $Y_p^{\rm S4}$ are model parameters of this particular case of CMB-S4 forecast. Notice that no BBN constraints are applied in this CMB-alone case to break the degeneracy between $N_{\rm eff}^{\rm S4}$ and $Y_p^{\rm S4}$ in advance. For upcoming surveys like CMB-S4, delensing will help to break their degeneracy and improve their constraints \cite{Green:2016cjr,Hotinli:2021umk}.} We also use the correlations of these three parameters from the same {\it Planck} MCMC chains for $N_\nu \ne 3$ adopted in the previous sections. Both cases are shown in Figure \ref{fig:futureYp} respectively as the dashdotted-red and the dashdotdotted-cyan curves. We again find
that combining BBN+CMB reduces the uncertainty
budget substantially, so that even at the
present $Y_p$ sensitivity, $\sigma(N_\nu) \approx 0.05$, very close to the neutrino heating limit, and substantially better than the CMB alone.
To push below the neutrino heating limit
is somewhat less demanding of \he4 observations,
albeit challenging, requiring
$\sigma(Y_p) < 0.001$ when $f_{\rm sky}^{\rm S4} = 0.3$. 

To further improve the joint $\sigma(N_\nu)$ sensitivity, the correlation between $N_{\rm eff}^{\rm S4}$ and $Y_p^{\rm S4}$ plays a crucial role. This correlation has impact on how the CMB likelihood convolves with the BBN likelihood on the $N_\nu$-$Y_p$ plane, determining the spread of the combined $N_\nu$ likelihood. In our toy model, we found that if the correlation coefficient $\rho^{\rm \ S4}_{(N_{\rm eff}, Y_p)} \sim -0.9$ at $f_{\rm sky}^{\rm S4} = 0.3$, {\it i.e.} highly anti-correlation between $N_{\rm eff}^{\rm S4}$ and $Y_p^{\rm S4}$, we can get a joint $\sigma(N_\nu) \sim 0.03$ ($\sim 1$ \% uncertainty) even with the
present $Y_p$ sensitivity.\footnote{For comparison, {\it Planck} 2018 has $\rho_{(N_{\rm eff}, Y_p)} \sim -0.67$. Moreover, the values of the other two correlation coefficients related to $\eta$ are insignificant for this $\sigma(N_\nu)$ forecast.} 
This agrees with forecasts in refs.~\cite{CMB-S4:2016ple,2019arXiv190704473A,Dvorkin2022}.
We note that attaining this precision requires
that the real CMB-S4 reaches
the key entries assumed in our toy model--not only the relevant parameter uncertainties, but also the strong $N_{\rm eff} -Y_p$ anticorrelation.

The orange curve in Figure \ref{fig:futureYp} shows the expected joint $\sigma(N_\nu)$ at different $Y_p$ precision using a CMB-S4 example likelihood forecast calculated and kindly provided by Benjamin Wallisch based on refs. \cite{Baumann:2015rya,Baumann:2017gkg,Wallisch:2018rzj,Hotinli:2021umk,Raghunathan}. This particular forecast, labeled as ``S4-approximate'' in the figure, is based on a set of internal-linear-combination noise curves derived from a version of the current design and a simple foreground model, and provides an approximation of what CMB-S4 is currently thought to look like. Given that the key inputs of this likelihood are
$\sigma(N_{\rm eff}^{\rm S4}) = 0.071$, $\sigma(Y_p^{\rm S4}) = 0.0045$, and $\rho^{\rm \ S4}_{(N_{\rm eff}, Y_p)} = -0.848$, our BBN+CMB-S4 forecast displays the joint $\sigma(N_\nu) \simeq 0.03$ for the most range of $\sigma(Y_p) > 0.001$ including the current observed \he4 uncertainty. Thus, it is very promising to probe the neutrino heating effect by combining current BBN constraints with the future CMB-S4 data.  

An expected CMB-S4 precision $\sigma(N_\nu) \simeq 0.03$ also reaches the Standard Model lower limit $\Delta N_\nu > 0.027$ for a scalar that was ever thermally populated (Eq.~(\ref{eq:dNlimSM}) with $g_{X,\rm eff}=1$).  
This will make CMB-S4 a powerful probe not only of Standard Model physics but also of departures from it.
Importantly, BBN plays a critical role here:
the CMB-only ``S4-approximate'' constraint only reaches $\sigma(N_\nu) \simeq 0.07$ if the BBN relation $Y_p(\eta)$ is not used
(long-dashed horizontal dashed line in Fig.~\ref{fig:futureYp}).

We also note that this forecasted S4 CMB-only limit $\sigma(N_\nu) \simeq 0.07$ would still be challenging to match with BBN alone.
To catch up with such a sensitivity as well in the BBN-only case, 
future $^4$He astronomical observations need to reach $\sigma(Y_p) \simeq 0.001$.
The lesson is that BBN and the CMB are most powerful when working together.

Finally, CMB-S4 should also improve the CMB measure of \he4 to $\sigma(Y_p^{\rm CMB}) \simeq 0.0045$.  While this will improve the precision by a factor of $\sim 4$ from the {\em Planck} value, it is still larger then the current errors $\sigma(Y_p^{\rm obs}) = 0.0033$ from astronomical observations.
Even so, the independent and clean $Y_p^{\rm CMB}$ measurement will offer
an important consistency check on $Y_{p, {\rm obs}}$
and its systematics.
Along similar lines, another independent cross-check would be measurement of $Y_p$ from helium signatures in the kinetic Sunyaev-Zel'dovich effect \cite{Hotinli:2022jna}.

\section{Discussion and Conclusions}
\label{sect:conclude}

BBN constraints on $N_\nu$ have long stood as a prototypical example
of cosmological probes of new physics.
Now CMB measurements have reached similar precision.
We find that these independent BBN and CMB 
measures of $N_\nu$ are consistent with each
other and with the Standard Model; this represents
a non-trivial success of hot big bang cosmology.
We have explored the consequences of using BBN and the CMB jointly and separately to look for new physics.
First, we updated the joint BBN+CMB constraints for the case where $N_\nu$ does not change
between these two epochs. The resulting limits
in Eqs.~(\ref{eq:DeltaNnu2side} - \ref{eq:DeltaNnu3})
tighten the constraints new physics in the early universe, which we illustrated with several examples.

Second,
advances in the BBN and CMB precision now also allow us to meaningfully probe {\em changes} in
$N_\nu$ and $\eta$ between these two epochs.
We find the the current data is consistent with the Standard Model, with only modest departures allowed. We applied our limits to
early dark energy models for the $H_0$ tension,
finding that BBN+CMB constraints can be significant.
We also placed bounds on changes in $\eta$ that limit
entropy production between BBN and the CMB.

It bears repeating that the limits we have set are valid when an effective value for
$N_\nu$ adequately captures the effect of new physics. Our results apply to situations
where the only important perturbation adds a radiation-like component to the cosmic energy density, and where there is no additional effects on structure growth as probed by the CMB.
Moreover, our work requires that $N_\nu$ takes a constant value throughout the BBN epoch,
and similarly for the CMB, though possibly with a different value.
Cases where our treatment is inadequate require dedicated analyses;
these include perturbations with density that does not
evolve as radiation such as early matter domination, or where interactions with the baryon-photon plasma can affect reaction rates.
And finally, we have assumed the primordial lithium problem does not have a new physics solution \cite{FO2022}; were that to be the case, a bespoke analysis is again required.

To further sharpen BBN and CMB constraints on new physics, we look forward to new precision 
results in 
nuclear experiments, nuclear theory, astronomical observations, and CMB measurements. BBN predictions for deuterium still have larger errors than D/H observations, due to the uncertainties in major deuterium destruction rates $d(d,n)\he3$, $d(d,p)t$, and $d(p,\gamma)\he3$ (listed by their contribution to the predicted D/H uncertainty). We point out that measurements of the $d+d$ reaction are the most urgently needed, and better theoretical understanding of these rates would also be illuminating.
Also, if BBN theory errors can be reduced via new cross section measurements, D/H can become a stronger probe of $N_\nu$.  Improvements in nuclear theory would be useful as well, particularly for deuterium rates: ref.~\cite{Pitrou2021} use nuclear theory in deriving thermonuclear rates, and find D/H abundances that are discrepant with the results of the empirical approach adopted here and in ref.~\cite{Pisanti2021}.

Improved astronomical measurements of the light elements can also advance the field.
For D/H, it may be challenging to improve the
already small error in the mean, but to continue to test for systematics, it remains critical to find additional good systems across as wide as possible a range of redshift and metallicity.
As we have shown, improvements in astronomical \he4 observations will sharpen $N_\nu^{\rm BBN}$ and thus 
sharpen the comparison with $N_\nu^{\rm CMB}$.
And of course, an unambiguous solution to the lithium problem remains outstanding
(but see ref.~\cite{FO2022}).

Finally, the upcoming CMB-S4 will improve $N_{\rm eff}$ measurements. The projected CMB-S4-alone precision $\sigma(N_{\rm eff}) \simeq 0.07$ will then likely provide a stronger probe of $N_\nu$ than BBN. To catch up with such CMB-S4 precision on the BBN side, future primordial $^4$He observations must reach $\sigma(Y_{p, {\rm obs}}) \simeq 0.001$, $\sim 1/3$ of the current (already-excellent) uncertainty. Moreover, the predicted BBN+CMB-S4 joint precision $\sigma(N_\nu) \simeq 0.03$ bring in reach new regimes in both conventional and exotic physics.
We will be able to discriminate for the first time
the difference $N_{\rm eff} - 3 = 0.044$ for the effect of neutrino heating.
And this sensitivity to $N_\nu$ can reveal the effects of new particles that were in thermal equilibrium at temperatures above the masses of all Standard Model particles.
We look forward to this view of the early Universe coming into focus.

\acknowledgments
We are grateful to Benjamin Wallish for sharing simulation results for CMB-S4 projections.
BDF is pleased to thank Cora Dvorkin and Laura Marcucci for illuminating discussions.  We are also grateful for discussions with many participants in the workshop on Latest Advances in the Physics of BBN and Neutrino Decoupling.
The work of K.A.O.~is supported in part by DOE grant DE-SC0011842  at the University of
Minnesota.  J.S.~is supported in part by DOE CAREER grant DE-SC0017840.

\bibliography{BBN_Nnu.bib}
\bibliographystyle{JHEP.bst}

\end{document}